\documentclass{JFM-FLM_Au}
\usepackage{amsmath,amssymb,bm}
\usepackage{booktabs}
\usepackage{array}
\newcolumntype{P}[1]{>{\raggedright\arraybackslash}p{#1}}
\usepackage{xcolor}
\usepackage{etoolbox}
\makeatletter
\patchcmd{\@maketitle}{(Received xx; revised xx; accepted xx)\hfill}{}{}
  {\PackageError{manuscript}{Could not remove template dates}{}}
\patchcmd{\@maketitle}{\addvspace{4.6pc}}{\addvspace{1pc}}{}
  {\PackageError{manuscript}{Could not adjust template banner space}{}}
\def\ps@titlepage{%
  \def\@oddhead{\@underjournal\hfil}%
  \let\@evenhead\@oddhead
  \def\@oddfoot{\hfil\thepage}%
  \def\@evenfoot{\thepage\hfil}}
\apptocmd{\ps@headings}{%
  \def\@oddfoot{\hfil\thepage}%
  \def\@evenfoot{\thepage\hfil}}{}{}
\AtBeginDocument{\pagestyle{headings}}
\makeatother

\newcommand{\figpdf}[1]{\IfFileExists{#1}{\includegraphics{#1}}%
  {\fbox{\texttt{[pending figure]}}}}

\newcommand{\clw}[1][]{c_{\mathrm{lw}#1}}  %

\lefttitle{Q. Li}
\righttitle{Selection versus residuals for the DJL equation}

\title{Selection versus residuals: deep-learning solvers for the
Dubreil-Jacotin--Long equation and application to mode-2 internal
solitary waves}

\author{Qiang Li\aff{1} and Yaoyin Zhang\aff{1}}
\affiliation{\aff{1}Shenzhen Key Laboratory for Marine IntelliSensing and Computation, Shenzhen International Graduate School, Tsinghua University, China}
\corresau{Qiang Li, \email{li.qiang@sz.tsinghua.edu.cn}}

\begin{document}
\maketitle

\begin{abstract}
Neural solvers offer structural flexibility for the Dubreil-Jacotin--Long
(DJL) equation of internal solitary waves, which also admits the trivial state, oscillatory wavetrains and
broadened near-solutions. A solver must select the physical wave before resolving it. Turkington's variational principle selects the
wave through the nonlinearity of the available potential energy (APE).
An exact virial partition shows that every localized wave carries more
kinetic energy than APE, so a quadratic APE would favour a broadened
field. The nonlinear APE enhancement, $1.27$--$2.06\times$ for the computed
waves, reverses this ordering, although the preferred width remains
weakly selected. From a shared KdV-derived start, a constrained Deep Ritz objective
recovers the mode-1
wave, whereas strong- and weak-form residual objectives, which vanish at every exact solution and therefore do not favour the physical wave over other roots, lose the prescribed APE within one
descent round. The recovered wave is under-resolved because its
residual is dominated by high-wavenumber error that the energy barely
penalizes, and a residual refinement at exact APE then meets every
acceptance criterion. Structurally excluding mode 1 from the trial space, we compute
mode-2 waves on four stratifications. A computed wave core qualitatively reproduces the
two-lobe structure observed on the South
China Sea shelf. In the deep basin, the selected core misses the
residual criterion, and residual refinement widens it with little change in the energy objective. The two resolved linear estimates suggest weaker relative mode-1
radiation at larger ratios of wave half-width to resonant wavelength.
\end{abstract}

\section{Introduction}\label{sec:intro}

Internal solitary waves are among the most energetic coherent
structures of the stratified ocean. They form where tidal flow meets
topography in a stratified water column, carry energy and momentum
across whole basins, and drive mixing and sediment resuspension on
continental shelves \citep{helfrich2006}. The South China Sea, the
setting of the final computations in this paper, hosts the largest
examples on record. Transbasin internal tides generated from Luzon Strait
steepen into solitary waves whose displacements exceed a hundred
metres, and their formation and fate have been traced across the basin
by coordinated observational programmes \citep{farmer2011,li_jpo2011}. Most of
that record concerns mode-1 waves, which displace the pycnocline in one
direction. Mode-2 waves, double-lobed disturbances that deform the
pycnocline about an interior node and travel more slowly, are firmly
established as well, and their observational record extends from the
laboratory experiments of \citet{davis1967} to moored time series. On
the continental slope of the northern South China Sea, mode-2 solitary waves with
isopycnal displacements of tens of metres arrive preferentially in
winter, when mode-1 waves are rarely observed \citep{yang2009}. On the New
Jersey shelf they appear in summer as short-lived trailing components
of the nonlinear internal wave field, radiating a visible mode-1 tail
\citep{shroyer2010}. That contrast poses a quantitative question. A
mode-2 wave can lose energy through resonant radiation into mode 1. How much energy is lost over a given propagation distance, and what controls that loss?

The theory of these waves at observed amplitudes is fully nonlinear.
Weakly nonlinear descriptions of Korteweg--de~Vries (KdV) type are
indispensable for scaling insight but degrade quantitatively at
oceanic steepness. The steady description of record is the
Dubreil-Jacotin--Long (DJL) equation, introduced by
\citet{dubreil1934} and \citet{long1953}. It is a nonlinear elliptic
eigenvalue problem, exact within the steady Boussinesq framework, and
its finite-amplitude solutions differ strongly from their weakly
nonlinear counterparts \citep{stastna2002}. Solving the DJL equation
accurately for prescribed continuous stratifications is therefore the
gateway computation for internal solitary-wave physics. Classical
solvers are mature. \citet{turkington1991} resolve the question
of which among its many solutions is the wave by characterizing the
physical state variationally, as the minimizer of a displacement-gradient
functional at fixed available potential energy (APE), and they give a
constructive iteration for that problem in which each step is a linear
elliptic solve. \citet{stastna2002} extend the computations to large
amplitude and background current, and the pseudospectral continuation
of \citet{dunphy2011}, which underlies the open-source solver
\textsc{PyDJL}, makes reference-quality solutions available in seconds
per wave. Fully nonlinear simulations also show that an internal solitary wave
carries more kinetic energy than APE \citep{lamb2009}. This paper leans
on that maturity.
The classical solvers supply its references, and nothing below competes
with them on per-wave speed or accuracy. What they do not offer is
structural flexibility. They are per-wave algorithms on fixed meshes,
and sectors of solution space carved out by symmetry or projection
require changes to their trial spaces or iterations. A neural trial
space accepts such restrictions by construction, and the mode-2
computations of this paper rest on that property.

Deep-learning solvers for partial differential equations come in three
recognized families. Physics-informed neural networks (PINNs) minimize the strong
residual of the equation at collocation points \citep{raissi2019},
weak-form methods (variational-PINN) test the residual against a basis of test functions
\citep{kharazmi2021}, and Deep Ritz methods minimize a variational
functional \citep{eyu2018}. The DJL equation puts a sharp question to
all three, because it does not by itself select the solitary wave
among the trivial state, oscillatory wavetrains and broadened
near-solutions that also satisfy it. A solver must therefore perform
two distinct tasks. It must
\emph{select} the solitary wave among these candidates, and it must
\emph{resolve} the selected wave to a prescribed residual. Classical
algorithms inherit selection from the fixed-APE structure of
Turkington's iteration and never face the difficulty. A neural solver
is defined by the objective its user chooses, and the choice decides
whether it inherits a selection principle. A non-negative residual
objective vanishes at every exact root and so supplies no energetic
ranking among them, whereas a variational objective does. The
literature on eigenvalue problems already records the symptom. For
linear eigenvalue problems, residual-based networks need an added
normalization term to avoid the trivial eigenfunction \citep{jin2022}, and
imbalance between the gradients of the terms of a composite objective
is a recognized source of failure in physics-informed optimization
\citep{wang2021}. For solitary waves of nonlinear wave equations a
two-stage procedure that fits an initial guess before minimizing the
residual has proved effective \citep{song2024}. The DJL problem
combines these difficulties with a nonlinear eigenvalue, a constraint
that is itself nonlinear in the displacement, and competing
near-solutions that a normalization alone does not exclude.

This paper therefore asks three questions. In the physics of the
constrained problem, what makes the solitary wave, rather than a broadened
field, the minimizer of Turkington's functional? When the same neural
trial space, constraint treatment and initial field are used with a
variational objective and with residual objectives, which of them
selects the wave, and why? Once the wave is selected, what limits its
resolution, and how can that limit be removed? The answers bear
directly on mode-2 waves, for which selection is harder because the
unrestricted variational problem favours mode 1, and on the radiation
of those waves into mode 1, which is the oceanographic question above.

The paper proceeds as follows. Section~\ref{sec:djl} states the DJL
equation and Turkington's principle, derives two exact identities that
connect the principle to the energetics of the wave, and examines the
constrained energy landscape along families of deformations of the
wave. Section~\ref{sec:formulations} places the strong-form, weak-form
and constrained Deep Ritz objectives in one trial space and states in
advance what each is expected to do. Section~\ref{sec:verification}
sets out the acceptance criteria and compares the three objectives from
a common start, from reference fields and from random fields.
Section~\ref{sec:mode1} turns from selection to resolution.
Section~\ref{sec:mode2} excludes mode 1 from the trial space, computes
mode-2 waves and quasi-steady cores on four stratifications ending with
a South China Sea profile, and estimates their radiation into mode 1.
Section~\ref{sec:discussion} draws the conclusions together with their
limitations, and the appendices give the numerical details and two
supplementary computations.

\section{The DJL equation and its variational characterization}
\label{sec:djl}

\subsection{Governing equation and the selection problem}
\label{sec:djl:eq}

We consider two-dimensional, inviscid, incompressible, non-rotating
Boussinesq flow with a stable continuous ambient stratification, no
background current, a flat impermeable bottom and a rigid lid. The
nondimensional frame rests on three scales. With $H$ the depth,
$\Delta\rho$ the top-to-bottom density difference and
$g'=g\,\Delta\rho/\rho_0$ the reduced gravity, lengths, including the
isopycnal displacement $\eta$, are measured in units of $H$,
velocities in units of $\sqrt{g'H}$ and the squared buoyancy frequency
in units of $g'/H$, so that dimensional speeds and stratifications are
recovered as $c^{*}=\sqrt{g'H}\,c$ and $N^{*2}=(g'/H)N^2$. The fluid
occupies $0\le z\le1$. The ambient stratification enters through the
scaled density $\bar s(z)$, with $\bar s(0)=1$ and $\bar s(1)=0$, and
\begin{equation}
  N^2(z)=-\frac{\mathrm{d}\bar s}{\mathrm{d}z},\qquad
  \int_0^1N^2\,\mathrm{d}z=1 ,
  \label{eq:N2}
\end{equation}
in which the unit integral is not an extra assumption but the scaling
itself, since it restates that the full density difference has been
absorbed into $g'$. A wave travelling steadily at speed $c$ through
fluid at rest at infinity, symmetric about $x=0$, has an isopycnal
displacement that satisfies the DJL equation \citep{stastna2002},
\begin{equation}
  \nabla^2\eta+\frac{N^2(z-\eta)}{c^2}\,\eta=0,\qquad
  \eta(x,0)=\eta(x,1)=0,\qquad \eta\to0\ \text{as}\ |x|\to\infty .
  \label{eq:djl}
\end{equation}
This is a nonlinear elliptic eigenvalue problem whose nonlinearity
enters solely through the evaluation of $N^2$ at the displaced height
$z-\eta$ of the fluid that now sits at $z$. Mode-1 solitary waves
travel faster than the linear long-wave speed $\clw$ of the
stratification, whereas higher modes travel below it, and at the
amplitudes observed in the coastal ocean the solutions depart strongly
from their weakly nonlinear descriptions \citep{stastna2002}. The fully
nonlinear equation, rather than an amplitude expansion, is therefore
the object solved throughout. Computations impose $\eta(L,z)=0$ on a
truncated domain, and physical admissibility is a separate question.
In the non-overturning class considered here the map to upstream
height must satisfy $0\le z-\eta\le1$ and $1-\eta_z>0$, which a field
that meets the boundary conditions need not do.

The selection problem that organizes this paper is already visible in
\eqref{eq:djl}, before any numerical method is chosen. The trivial
state $\eta\equiv0$ satisfies the equation at every speed. Periodic
wavetrains satisfy it at speeds below $\clw$, and on a truncated domain
they meet every boundary condition whenever a node falls at $x=L$.
Beyond these exact solutions, horizontally broadened quasi-linear
fields carry any prescribed APE while their residual tends to zero as
width and domain grow together. These near-solutions differ from the
wave in shape rather than in how well they satisfy the equation. The
numerical record of this study contains endpoints of all three kinds.
The wave speed, moreover, is not a datum but an eigenvalue. A method
for \eqref{eq:djl} therefore answers two distinct questions. One is
how accurately it can represent a solution. The other, logically
prior, is which solution it converges to, and the classical resolution
of that question is variational.

\subsection{Turkington's variational principle and two exact identities}
\label{sec:djl:variational}

The physically realized mode-1 wave is characterized variationally
\citep{turkington1991} as the minimizer of the displacement-gradient
functional
\begin{equation}
  \mathcal{K}[\eta]=\tfrac12\iint_\Omega|\nabla\eta|^2\,
  \mathrm{d}x\,\mathrm{d}z,\qquad \Omega=[-L,L]\times[0,1],
  \label{eq:KE}
\end{equation}
over the fields that carry a prescribed APE $A>0$,
\begin{equation}
  \mathcal{F}[\eta]=\iint_\Omega f(z,\eta)\,\mathrm{d}x\,\mathrm{d}z=A,
  \qquad
  f(z,\eta)=\int_0^\eta\bigl[\bar s(z-\eta)-\bar s(z-\xi)\bigr]
  \mathrm{d}\xi\ \ge0 .
  \label{eq:APE}
\end{equation}
Since $\eta$ is even in $x$, every integral is evaluated on $[0,L]$
and doubled, so every energy and residual quoted below is a full-wave
value, and r.m.s. residuals are normalized by the domain measure. For
the $\tanh$ stratifications used here, the inner integral has a closed
form through the antiderivative of $\tanh$, verified against direct quadrature once per stratification. Differentiating \eqref{eq:APE} gives
\begin{equation}
  \frac{\delta\mathcal{F}}{\delta\eta}=\eta\,N^2(z-\eta),
  \qquad\text{hence}\qquad
  f(z,\eta)=\int_0^\eta\xi\,N^2(z-\xi)\,\mathrm{d}\xi .
  \label{eq:dFdeta}
\end{equation}
The second form follows from the first with $f(z,0)=0$, and it makes
the physics of the APE density explicit. The quantity $f$ is the work
done against buoyancy in carrying a parcel from its upstream height
$z-\eta$ to $z$, with $N^2$ evaluated at every height traversed. Its
quadratic approximation $\tfrac12N^2(z)\eta^2$ evaluates $N^2$ only at
the parcel's final position, and the difference between the two is
what selects the wave in \S\ref{sec:djl:landscape}.

Stationarity of the Lagrangian $\mathcal{K}-\lambda(\mathcal{F}-A)$
gives
\begin{equation}
  \nabla^2\eta+\lambda N^2(z-\eta)\eta=0,\qquad \lambda=\frac{1}{c^2},
  \label{eq:el}
\end{equation}
so the DJL equation is recovered with the wave speed entering as the
Lagrange multiplier of the APE constraint. The eigenvalue therefore
need not be treated as an unknown of the optimization. It is a
by-product of the constrained minimization, recovered a posteriori
from the Rayleigh identity that follows from multiplying \eqref{eq:el}
by $\eta$ and integrating by parts,
\begin{equation}
  c^2=\frac{\iint_\Omega N^2(z-\eta)\,\eta^2\,\mathrm{d}x\,\mathrm{d}z}
           {\iint_\Omega|\nabla\eta|^2\,\mathrm{d}x\,\mathrm{d}z} .
  \label{eq:rayleigh}
\end{equation}
A second exact consequence of stationarity follows from the
dilations $\eta(x/\beta,z)$, for which
$\mathcal{K}(\beta)=\tfrac12(I_x/\beta+\beta I_z)$ and
$\mathcal{F}(\beta)=\beta A$, with $I_x=\iint\eta_x^2$ and
$I_z=\iint\eta_z^2$. Stationarity at $\beta=1$ on the infinite domain
gives the Derrick-type (virial) identity
\begin{equation}
  \tfrac12(I_z-I_x)=\frac{A}{c^2} ,
  \label{eq:virial}
\end{equation}
whose boundary-corrected form on a truncated domain serves as an
acceptance check in \S\ref{sec:verification:certificates}. The
laboratory-frame velocities are $u=c\eta_z$ and $w=-c\eta_x$, so the
physical kinetic energy per unit span, in units of $\rho_0g'H^3$, is
$E_K=c^2\mathcal{K}$. Minimizing $\mathcal{K}$ at fixed APE is
therefore not the same as minimizing $E_K$ when $c$ varies, and every
energy ratio below refers to $\mathcal{K}$ unless stated otherwise.
Combining $E_K=\tfrac12c^2(I_x+I_z)$ with \eqref{eq:virial} gives an
exact energy partition,
\begin{equation}
  E_K-A=c^2I_x\ \ge0,\qquad
  \frac{I_z}{I_x+I_z}=\frac12\Bigl(1+\frac{A}{E_K}\Bigr) .
  \label{eq:partition}
\end{equation}
Since $u=c\eta_z$ and $w=-c\eta_x$, the kinetic energy splits into a
horizontal part $\tfrac12c^2I_z$ and a vertical part $\tfrac12c^2I_x$,
and \eqref{eq:virial} states that the APE equals their difference. A
localized solitary wave therefore carries more kinetic energy than
APE, the excess $c^2I_x$ being twice the kinetic energy of its
vertical motion, and equipartition is approached only in the long-wave
limit $I_x/I_z\to0$. The inequality is the one that \citet{lamb2009}
find in fully nonlinear simulations, where the excess reaches $30\%$
in quasi-two-layer stratifications, and \eqref{eq:partition} shows
that it holds for every localized DJL wave in the present setting. The
partition is the lever with which we interpret the selection margin
below.

\subsection{The energy landscape of the constrained problem}
\label{sec:djl:landscape}

All landscape and mode-1 results are computed on one benchmark
stratification, the single-$\tanh$ pycnocline
$\bar\rho(z)=1-\epsilon\tanh((z-z_0)/d)$ with $\epsilon=0.01$,
$z_0=0.8$ and $d=0.1$ used by \citet{dunphy2011}, for which
\[
  N^2(z)=\frac{b}{d}\,\mathrm{sech}^2\Bigl(\frac{z-z_0}{d}\Bigr),
  \qquad b=0.509 ,
\]
with $\epsilon$ entering only through the contrast
$\Delta\rho=0.0196$ that sets $g'$. We refer to it as the benchmark
profile. The prescribed APE spans the $10\times$ range $A=0.005$ to
$0.050$, with $A_{\mathrm{ref}}=0.0293$ as the reference value. This is
the APE of the original reference solution, whose crest displacement
is $0.21$ of the depth and whose $\max|u|/c=0.66$, a moderately
nonlinear wave well past the weakly nonlinear regime and well short of
overturning.

How strongly does the constrained minimization prefer the wave? The
broadening degeneracy noted in \S\ref{sec:djl:eq} threatens any
optimization-based method, so we measure the preference along the
deformations most relevant to it before any network enters. At
quadratic order in amplitude a horizontally broadened mode-1 state of
width $W$ at fixed APE has an $x$-gradient energy of order $A/W^2$ but
a $W$-independent $z$-gradient energy, so the existence of an interior
minimum in $W$ rests on non-quadratic structure. The audit takes
converged spectral reference waves $\eta^{\mathrm{ref}}$ at the five
amplitudes and deforms them at fixed APE in two ways. The horizontal
dilations
\begin{equation}
  \eta_\beta(x,z)=s(\beta)\,\eta^{\mathrm{ref}}(x/\beta,z)
  \label{eq:family}
\end{equation}
stretch the wave for $\beta>1$ and compress it for $\beta<1$, with the
amplitude factor $s(\beta)$ fixed by Newton iteration so that
$\mathcal{F}[\eta_\beta]=A$ to machine precision. Each dilated field is
evaluated on its own stretched domain $[0,\beta L]$, so its far field
is never truncated and $\beta\to\infty$ is an expanding-domain limit
rather than a trajectory on a fixed domain. Two vertical deformations,
$\eta^{\mathrm{ref}}(x,z\pm0.15\sin\pi z)$, move the crest down or up
and are rescaled in the same way. We map $\mathcal{K}$ along the
dilation family at 261 widths from $\beta=0.45$ to $120$, 41 of them
within $10\%$ of the wave, and check the audit itself against the
virial identity \eqref{eq:virial} and under quadrature refinement.

\begin{figure}
  \centerline{\figpdf{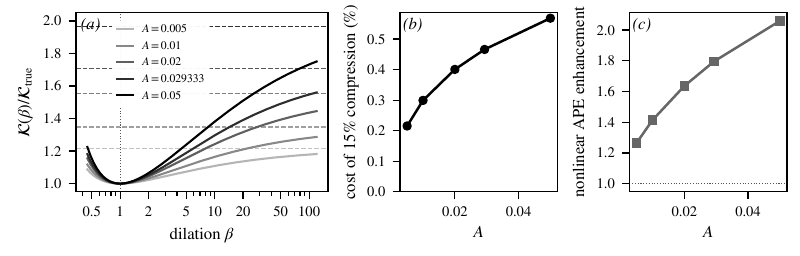}}
  \caption{The constrained energy landscape of the mode-1 wave.
  ($a$)~$\mathcal{K}$ along the APE-preserving dilation family at each
  tested amplitude, normalized by the value for the wave, for
  $0.45\le\beta\le120$. Dashed lines mark the broadened limits
  $AI_z/(Q\,\mathcal{K}_{\mathrm{true}})$. ($b$)~The increase of
  $\mathcal{K}$ under a $15\%$ compression ($\beta=0.85$) as a function
  of amplitude. ($c$)~The nonlinear APE enhancement $2A/Q$
  ($1.27$--$2.06$), which lifts the broadened limit above the wave.}
  \label{fig:dr0landscape}
\end{figure}

The constrained energy landscape is shown in figure~\ref{fig:dr0landscape}. At every amplitude
the minimum of $\mathcal{K}$ along the dilation family lies at the
sampled point $\beta=1$, with both neighbours at $\beta=1\pm0.005$
higher, so the true wave has the lowest $\mathcal{K}$ of any state
\emph{within the tested dilation and deformation family}. The audit
does not exclude states outside that family, which are left to the
acceptance criteria of \S\ref{sec:verification}. The preference is
strong against vertical deformations, which raise $\mathcal{K}$ by
$26\%$--$47\%$, and against large horizontal ones. As $\beta\to\infty$
the energy tends from below to a limit $22\%$--$97\%$ above the wave.
It is weak against small horizontal deformations. Compressing the wave
by $15\%$ ($\beta=0.85$) raises $\mathcal{K}$ by only
$0.22\%$--$0.57\%$ (panel $b$), least at the smallest amplitude and
about $1.6$ times the cost of an equal stretch. For the APE-preserving dilation, differentiation at the reference wave gives $\mathcal{K}'(1)=\tfrac12(I_z-I_x)-A/c^2=0$, in which the equality follows from the Rayleigh and virial identities. Thus, near $\beta=1$, the leading change in energy is quadratic,
\[
\frac{\mathcal{K}(\beta)}{\mathcal{K}_{\mathrm{true}}}
\approx 1+\tfrac12\kappa(\beta-1)^2.
\]
Fitting this expression to the computed energies near \(\beta=1\) gives \(\kappa=0.15\) at \(A=0.005\), increasing to \(0.39\) at \(A=0.05\).

Why does the wave beat its own broadened limit? As $\beta\to\infty$ at
fixed $A$ the amplitude falls as $\beta^{-1/2}$, $f$ becomes quadratic
and $\mathcal{F}\to\tfrac12s^2\beta Q$ with
$Q=\iint N^2(z)\,\eta_{\mathrm{true}}^2$. Hence
$\mathcal{K}_\infty=AI_z/Q$, and with \eqref{eq:partition}
\begin{equation}
  \frac{\mathcal{K}_\infty}{\mathcal{K}_{\mathrm{true}}}
  =\underbrace{\frac{2A}{Q}}_{\text{APE enhancement}}\times
   \underbrace{\frac{I_z}{I_x+I_z}}_{\tfrac12(1+A/E_K)}
  \label{eq:broadlimit}
\end{equation}
Thus, the two factors pull in opposite directions. The second factor is below unity
for every localized wave, because $E_K>A$ by \eqref{eq:partition}. The
computed waves carry $8\%$--$11\%$ more kinetic energy than APE, so
this factor sits $3.8\%$--$5.0\%$ below unity, and if the APE were
exactly quadratic the wave could always lower $\mathcal{K}$ by
spreading. The broadened limit would then be the preferred state, which
is the broadening degeneracy in variational form. The first factor
reverses the ordering. The APE of the wave is $1.27$--$2.06$ times its
own quadratic-order estimate $Q/2$. A broadened field at the same
$A$, being lower in amplitude and more nearly linear, forfeits that
enhancement and must pay the same APE with more gradient energy. The
path-integral form \eqref{eq:dFdeta} shows where the enhancement comes
from. At the crest of the reference wave the fluid displaced furthest
has come from the pycnocline centre, $z-\eta\approx0.80=z_0$, and now
sits near $z=0.59$, about two pycnocline thicknesses lower, where $N^2$
is about $6\%$ of its peak value. The quadratic estimate charges this
parcel only the local $N^2$, whereas the true APE density integrates
the strong stratification it has crossed. The enhancement sets the
height of the broadened limit, not the shape of the curve near
$\beta=1$. Both factors tend to unity in the weakly nonlinear limit, so the energy penalty for broadening, ${\mathcal{K}_\infty}/{\mathcal{K}_{\mathrm{true}}}-1$, tends to zero. Consistent with this weakening preference for localization, figure~\ref{fig:dr0landscape} shows that the energy penalties for broadening and compression are smallest at the lowest APE considered.

This weak local preference has a consequence for any
optimization-based solver. Widths differing by $|\beta-1|$ differ in
$\mathcal{K}$ by only $\tfrac12\kappa(\beta-1)^2$, so an objective
uncertain by a relative amount $\nu$ cannot rank widths within
$(2\nu/\kappa)^{1/2}$ of the wave, which is $0.23$--$0.52$ for an
illustrative $\nu=1\%$--$2\%$. This extrapolation of the local
quadratic fit is not an error bound, and correlated quadrature errors
can cancel in differences of objectives, but it shows why small
differences in $\mathcal{K}$ must be resolved, and it motivates the
deterministic quadrature and controlled descent of the deep-learning formulations.

\subsection{Classical algorithms and the role of a neural trial space}
\label{sec:djl:classical}

Two classical algorithm families realize the constrained minimization,
and both serve this paper as ancestry and as benchmarks.
\citet{turkington1991} give a constructive iteration that solves the
variational problem without any nonlinear optimization. The
stratification source is frozen at the current iterate, the resulting
linear elliptic problem is solved, the eigenvalue $\lambda=1/c^2$ is
updated through a closed-form quadrature ratio, and the field is
rescaled to restore the APE exactly, so that selection and control of
the amplitude are supplied by the fixed-APE structure of the iteration
itself. A finite-difference implementation of that iteration reaches a
dense-grid residual of $2.51\times10^{-4}$ on the benchmark problem in
about $22$~s per wave. The pseudospectral realization of
\citet{dunphy2011}, available as \textsc{PyDJL} \citep{pydjl}, a
Python translation of \textsc{DJLES} \citep{djles}, combines the same
iteration with spectral discretization and amplitude continuation. Its
waves, converged to a tolerance of $10^{-8}$, carry a dense-grid
residual four orders of magnitude below any acceptance threshold
applied to a network state, at about $2$~s per wave on a $256^2$ grid,
and they supply every reference solution in this paper
(table~\ref{tab:benchmarks}).

Both algorithms get selection exactly right, and that is the standard
against which \S\ref{sec:formulations} measures the learning
formulations. What they do not do, as implemented and as ordinarily
run, is compute mode 2. Both initialize from the leading linear
eigenmode. On a stratification symmetric about mid-depth the
antisymmetric sector is invariant under the Turkington iteration, so a
mode-2 wave can be computed with an unmodified mode-1 solver on the
upper half of the column. On the asymmetric stratifications of the
ocean no such sector exists, and the deeper obstacle is the selection
principle itself. The mode-1 wave is
the global minimizer of $\mathcal{K}$ at prescribed APE, and on the
stratifications of this paper a mode-2 state carries markedly more
$\mathcal{K}$ at equal APE, a gap quantified in \S\ref{sec:mode2}, so
a mode-2 initial field alone does not remove the preference for mode 1.
A global preference guarantees neither convergence from every initial
field nor departure from an invariant sector, and no classical
random-start success rate is inferred here. Computing mode 2 from the
variational principle therefore requires restricting the search space
so that the mode-1 branch is excluded from it. Such restrictions can be
imposed in classical discretizations, by a half-domain solve, an
adapted basis or a projected iteration, but they are natural in a
representation that accepts them by construction, and that is the role
of the neural trial space in this paper. It is not a faster or more
accurate solver for individual mode-1 waves.

\section{Three deep-learning formulations}\label{sec:formulations}

\subsection{What the formulations share}
\label{sec:formulations:shared}

\begin{figure}
  \centerline{\figpdf{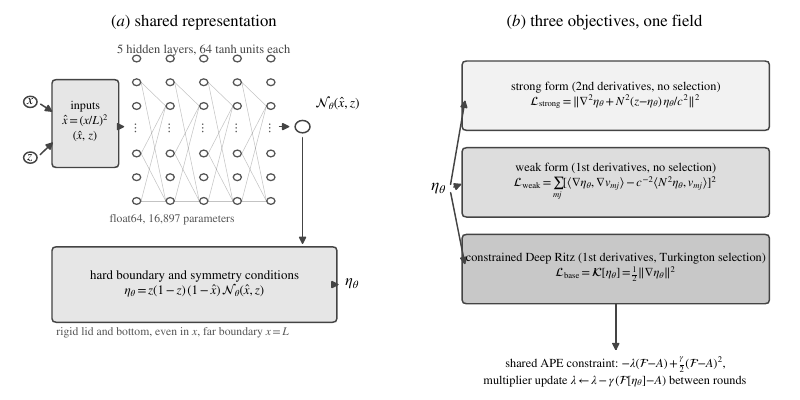}}
  \caption{The shared representation and the three objectives.
  ($a$)~The inputs $(\hat x,z)$, with $\hat x=(x/L)^2$, pass through
  five hidden layers of 64 $\tanh$ units (16,897 parameters), and the
  ansatz \eqref{eq:ansatz} enforces every boundary and symmetry
  condition exactly. ($b$)~The three objectives evaluated on the same
  field $\eta_\theta$. The strong form squares the pointwise residual
  of \eqref{eq:djl} and involves second derivatives. The weak form
  squares its projection onto a fixed test basis after one integration
  by parts and involves first derivatives. The constrained Deep Ritz
  objective is $\mathcal{K}$ under the APE constraint \eqref{eq:al},
  involves first derivatives and carries the selection principle of
  \S\ref{sec:djl:variational}. ``No selection'' denotes the absence of
  an energetic ranking among exact roots, not an inability to retain a
  root.}
  \label{fig:architecture}
\end{figure}

Three formulations of \eqref{eq:djl} as a learning problem are implemented and compared in this paper. Figure~\ref{fig:architecture} fixes what they share and where they differ. All three represent the
displacement by the same network inside the same ansatz, evaluate it in
double precision on the same fixed quadrature grid, and impose the APE
constraint through the same augmented Lagrangian. In the decisive
comparison of \S\ref{sec:verification:comparison} they also start from
the same fitted initial field. They diverge in the objective, so the
shared elements are specified once, here, and the constants are
collected in Appendix~\ref{app:numerics}.

The displacement is represented by a fully connected network
$\mathcal{N}_\theta$ with $\tanh$ activations, wrapped in an ansatz that
enforces every boundary and symmetry condition of the truncated problem
structurally rather than through penalty terms,
\begin{equation}
  \eta_\theta(x,z)=z(1-z)\,(1-\hat x)\,\mathcal{N}_\theta(\hat x,z),
  \qquad \hat x=(x/L)^2 .
  \label{eq:ansatz}
\end{equation}
The factor $z(1-z)$ enforces the conditions at bottom and lid, the even
input $\hat x$ renders $\eta_\theta$ symmetric about $x=0$ with
$\partial_x\eta_\theta(0,z)=0$ identically, and $(1-\hat x)$ pins
$\eta_\theta(L,z)=0$ while the network supplies the exponential decay in
the interior. The domain is adequate when $\mu_{\mathrm{tail}}(c)L\ge8$,
where $\mu_{\mathrm{tail}}$ is the slowest linear far-field decay rate
at the current speed, and the condition is rechecked whenever $c$
changes because $\mu_{\mathrm{tail}}\to0$ as $c\to\clw^{+}$. Physical
admissibility is not enforced by \eqref{eq:ansatz} and is checked
separately.

The APE constraint is imposed through the augmented Lagrangian
\begin{equation}
  \mathcal{L}_\gamma(\theta,\lambda)=\mathcal{L}_{\mathrm{base}}[\eta_\theta]
  -\lambda\bigl(\mathcal{F}[\eta_\theta]-A\bigr)
  +\frac{\gamma}{2}\bigl(\mathcal{F}[\eta_\theta]-A\bigr)^2 ,
  \label{eq:al}
\end{equation}
minimized over $\theta$ at fixed $(\lambda,\gamma)$ by L-BFGS with a
strong-Wolfe line search \citep{liu1989}. Adam \citep{kingma2015} is
used only in the preceding least-squares fit of the network to the
initial field (Appendix~\ref{app:numerics}). Between optimization rounds
the multiplier is updated as
$\lambda\leftarrow\lambda-\gamma(\mathcal{F}-A)$, and $\gamma$ is
increased only when the constraint violation fails to contract. The
multiplier is initialized for every objective at the value the initial
field implies through the Rayleigh quotient \eqref{eq:rayleigh},
$\lambda_0=1/c_0^2$. All functionals, the multiplier update and every
in-loop diagnostic are evaluated on a fixed tensor-product quadrature
grid of at least $512\times256$ nodes, with derivatives of
$\eta_\theta$ obtained by automatic differentiation and the
displaced-height argument of $N^2$ kept inside the differentiation
graph. No stochastic sampling is used. The weak width preference of
\S\ref{sec:djl:landscape} motivates resolving small differences of
$\mathcal{K}$ deterministically, and the fixed grid resolves them
(Appendix~\ref{app:numerics}). This is a choice made for this problem,
not a claim that stochastic optimization must fail. Descent is monotone within each round at fixed
$(\lambda,\gamma)$.  The
choice also has a measured basis. A tested configuration that uses Adam for the six rounds of
\eqref{eq:al} before the same L-BFGS restarts carries the state away from the wave and ends on a
broadened off-axis state with an objective $7.05\%$ above that of its
own start at the same final multipliers, whereas line-search L-BFGS on
the otherwise identical protocol converges onto the wave. Physical
checks of speed ($c\ge1.01\,\clw$), domain adequacy, crest position and
amplitude stop a Ritz or reference-initialized computation when
violated. In residual-objective computations from KdV-derived starts,
they are recorded only, so that the trajectory can be observed.

\subsection{The three objectives and their multipliers}
\label{sec:formulations:strong}
\label{sec:formulations:vpinn}
\label{sec:formulations:deepritz}

The \emph{strong form} (PINN), in the physics-informed lineage of
\citet{raissi2019}, substitutes the represented field into
\eqref{eq:djl} and minimizes the mean-square collocation residual on the quadrature grid
\begin{equation}
  r=\nabla^2\eta_\theta+\frac{N^2(z-\eta_\theta)\eta_\theta}{c^2},\qquad \mathcal{L}_{\mathrm{strong}}[\eta_\theta]=\frac{1}{|\Omega|}
  \iint_\Omega r^2\,\mathrm{d}x\,\mathrm{d}z
  \label{eq:lstrong}
\end{equation}
The mean-square residual $\mathcal{L}_{\mathrm{strong}}$ is also the accuracy measure used throughout. The implemented base objective is the unnormalized integral $B_{\mathrm s}=\iint_\Omega r^2\,\mathrm{d}x\,\mathrm{d}z=|\Omega|\,\mathcal{L}_{\mathrm{strong}}$, with $|\Omega|=16$ at $L=8$. The speed is frozen
within each optimization round and refreshed between rounds from the Rayleigh identity \eqref{eq:rayleigh}. This relation holds exactly for a DJL solution, but for an approximate field it generally differs from the speed that minimizes the squared residual, so its speed update is not itself a residual minimization step.

The \emph{weak form}, in the lineage of \citet{kharazmi2021}, tests the
same residual against $L^2$-orthonormal functions
$v_{mj}=\chi_j(x)\zeta_m(z)$ on $D=[0,L]\times[0,1]$, with
$\zeta_m=\sqrt2\sin(m\upi z)$ and
$\chi_j=\sqrt{2/L}\cos((j-\tfrac12)\upi x/L)$, after one integration by
parts,
\begin{equation}
  \mathcal{R}_{mj}=\iint_\Omega\Bigl(\nabla\eta_\theta\cdot\nabla v_{mj}
  -\frac{N^2(z-\eta_\theta)}{c^2}\eta_\theta v_{mj}\Bigr)
  \mathrm{d}x\,\mathrm{d}z,\qquad
  \mathcal{L}_{\mathrm{weak}}[\eta_\theta]=B_{\mathrm w}=\sum_{m,j}\mathcal{R}_{mj}^2 .
  \label{eq:weakres}
\end{equation}
All four boundary terms vanish
exactly, through the test functions at the walls and at $x=L$ and
through the even input at $x=0$, and the identity between the weak and
the projected strong residual holds to $4.2\times10^{-5}$ on the
quadrature grid with $h^2$ convergence under refinement. The base space
has $8\times16=128$ functions and resolves the spectral support of the
wave with a margin of a factor of several in each direction. A single
enlargement, fixed in advance, doubles both counts to 512. An
element-based test space of the $hp$-variational type
\citep{kharazmi2021} is also tested. The weak form is the intermediate
case of the comparison. It shares the derivative order of the
constrained Deep Ritz objective and the objective class of the strong
form, because \eqref{eq:weakres} vanishes at every root of \eqref{eq:el}
for any choice of test functions and carries no selection principle.

The \emph{constrained Deep Ritz} objective \citep{eyu2018} is
Turkington's functional itself,
\begin{equation}
  \mathcal{L}_{\mathrm{base}}[\eta_\theta]=\mathcal{K}[\eta_\theta],
  \label{eq:lritz}
\end{equation}
so that \eqref{eq:al} is exactly the augmented Lagrangian of the
variational characterization in \S\ref{sec:djl:variational}, and the
selection principle that the classical algorithms inherit by
construction is embedded in the objective. The wave speed is determined indirectly by the constrained minimization. At an exact stationary solution, the APE multiplier satisfies $\lambda=1/c^2$. In the computation, $c$ is evaluated from the resulting field using the Rayleigh identity \eqref{eq:rayleigh}.

The three objectives thus differ in two independent properties, the
derivative order and the presence of a selection principle, and the
weak form separates them. The multipliers make the difference precise.
At the wave the Ritz multiplier is $\lambda=1/c^2$ by \eqref{eq:el}, so
that with $E_K=c^2\mathcal{K}$
\begin{equation}
  \lambda A=\frac{A}{E_K}\,\mathcal{K}.
  \label{eq:selfscale}
\end{equation}
For the computed benchmark waves, $A/E_K\approx0.90$--$0.93$, with the inequality $A/E_K<1$ following from \eqref{eq:partition}.
The constraint term of the Ritz objective is therefore automatically
commensurate with the objective it constrains. A residual objective is
zero and stationary at every solution, so at an exact feasible root its
equilibrium multiplier is zero, provided $\nabla_\theta\mathcal{F}\ne0$.
A Rayleigh-initialized $\lambda_0$ is therefore not a neutral choice
for a residual objective, and the scale of $B_{\mathrm s}$ or
$B_{\mathrm w}$ away from a root, set by the quality of the field and
by the normalization, bears no relation to the constraint.

The expectations for the comparison are given in table~\ref{tab:taxonomy}, fixed before any numerical solution is measured. The residual objectives
are expected to fail from these starts and the Ritz objective to
recover the wave, and a weak-form recovery is designated a
counterexample to that empirical expectation, though not to the
mathematical absence of a ranking. From reference-initialized fields,
direct residual minimization is expected to reach a lower full
residual than the Ritz objective, and a larger test space to make more
of the residual visible, without either expectation implying an
ordering of nonconvex optimizer endpoints. Wave structure, feasibility
and the physical checks must accompany residual accuracy throughout,
since a collapsed field can have a small residual without carrying the
prescribed APE. 

\subsection{KdV-derived initial fields}
\label{sec:formulations:kdv}

Initial fields that require no DJL solution are built from KdV theory.
The long-wave eigenproblem
\begin{equation}
  \phi_{zz}+\frac{N^2(z)}{\clw^2}\phi=0,\qquad \phi(0)=\phi(1)=0,
  \label{eq:lw}
\end{equation}
furnishes $\clw$ and the mode-1 structure $\phi$, normalized to $+1$ at its largest magnitude, and the KdV coefficients \citep{dunphy2011}
\begin{equation}
  r_{10}=-\frac{3}{4E}\int_0^1(\phi')^3\mathrm{d}z,\qquad
  r_{01}=-\frac{\clw}{2E}\int_0^1\phi^2\mathrm{d}z,\qquad
  E=\int_0^1(\phi')^2\mathrm{d}z
  \label{eq:kdv}
\end{equation}
determine the solitary wave
\begin{equation}
  \eta_{\mathrm{KdV}}=-b_0\,\mathrm{sech}^2(x/\ell)\,\phi(z),\qquad
  b_0\ell^2=-\frac{6r_{01}}{\clw r_{10}},\qquad
  c_{\mathrm{KdV}}=\clw\Bigl(1+\tfrac23r_{10}b_0\Bigr),
  \label{eq:kdvstart}
\end{equation}
with $\mathrm{sign}\,b_0=\mathrm{sign}\,r_{10}$ and half-width
$\ell\operatorname{arcosh}\sqrt2$. The parameters $b_0$, $\ell$ and
$c_{\mathrm{KdV}}$ are rebuilt together to match the full nonlinear
APE, and the network is fitted to the resulting field in the
least-squares sense before the objective \eqref{eq:al} is activated.
Within the amplitude range of the construction, $|b_0|\le0.5$, the
KdV wave carries at most $\mathcal{F}=0.040$, so at $A=0.050$ the wave is
reached by continuation from smaller amplitudes instead
(\S\ref{sec:mode1:reachability}).

\section{Acceptance criteria and the measured comparison}
\label{sec:verification}

\subsection{Acceptance criteria and reference solutions}
\label{sec:verification:certificates}
\label{sec:verification:references}

A low value of the objective is not by itself evidence that the
physical wave has been found. Constraint-satisfying spurious states
exist, and in an exploratory strong-residual computation with
collocation sampling one state carries a residual below $10^{-3}$ while its APE
has fallen to about $10^{-7}$ of the prescribed value. A computed state
is therefore accepted only on a joint set of criteria evaluated a
posteriori on an independent dense grid, never on the quadrature grid
of the optimization. For mode 1 the criteria are the r.m.s. residual
of the full second-order equation \eqref{eq:djl} below $10^{-3}$, which
is the accuracy criterion, together with the crest position against
the reference, the bound $\max|u|/c=\max|\eta_z|<1$ with the correct
polarity, a relative APE error below $10^{-2}$, and the Rayleigh speed
within $10^{-3}$ of the independent spectral reference
(table~\ref{tab:certificates}). The asymmetry between optimization and
acceptance is deliberate. The weak and Ritz objectives contain first
derivatives only, while the residual criterion tests second-derivative
accuracy, so acceptance is an independent test rather than a
restatement of the objective. Because $c$ is a Rayleigh quotient it is
nearly stationary with respect to field error at the solution, so a small error in wave speed does not necessarily imply an equally accurate displacement field. We therefore report the maximum relative field error alongside the relative speed error.

The reference waves are computed with the pseudospectral solver of
\citet{dunphy2011}. Their dense-grid DJL residual is of order
$10^{-8}$, four orders below the acceptance threshold, and they enter
only validation and the reference-initialized diagnostics, never the
construction or optimization of KdV-derived starts. Two exact
identities supplement them. The Rayleigh identity recomputes $c$ from
the field alone. The virial identity \eqref{eq:virial}, corrected for
domain truncation by the boundary term that the dilation family
generates at $x=L$, reads
\[
  \tfrac12(I_z-I_x)+L\int_0^1\eta_x(L,z)^2\,\mathrm{d}z=\mathcal{F}/c^2 ,
\]
with $I_x$, $I_z$ and $\mathcal{F}$ the full-wave integrals. On the
15 recorded mode-1 solitary waves it holds to between
$5\times10^{-6}$ and $3.3\times10^{-4}$, and it catches spurious states
that the residual band misses. The displaced-broadened state of
Appendix~\ref{app:uniqueness} passes its residual band while violating
the virial identity at $6.1\times10^{-2}$, about $200$ times the
largest value on a true wave.

\begin{table}
\centering
\caption{Acceptance criteria for the computed waves. For mode 1, the
full residual measures equation accuracy, whereas items 2--7 identify
the solitary-wave structure. For asymmetric mode-2 cores, residual
accuracy is assessed through the orthogonal remainder $r_\perp$, with
the full residual and the mode-1 forcing $f_1$ reported separately. In
the antisymmetric sector of a symmetric stratification, $f_1\equiv0$
and the full homogeneous-DJL residual is used.}
\label{tab:certificates}
\footnotesize
\begin{tabular}{@{}c P{7.2cm} P{3.4cm}@{}}
\toprule
Item & Mode-1 acceptance criteria & Threshold \\
\midrule
1 & res\textsubscript{rms} of the full DJL equation & $\le 10^{-3}$ \\
2 & crest $z$ against the crest height of the reference solution & $\pm 0.05$ \\
3 & crest $x$ on the symmetry axis & $|x_{\rm crest}| \le 0.05$ \\
4 & $\max|u|/c = \max|\eta_z|$ and displacement polarity & $< 1$, correct polarity \\
5 & APE constraint $|F(\eta)-A|/A$ & $< 10^{-2}$ \\
6 & $c$ relative error against the spectral reference & $< 10^{-3}$ \\
7 & $\eta$ overlay against the reference & figure \\
\bottomrule
\end{tabular}

\vspace{6pt}
\begin{tabular}{@{}c P{7.2cm} P{3.4cm}@{}}
\toprule
Item & Mode-2 acceptance criteria & Threshold \\
\midrule
1 & checkpoint-level stabilization, final restart $d_{\rm rel}$ &
$\le 10^{-6}$ \\
2 & constraint $|F-A|/A$ & $< 10^{-2}$ \\
3\textsuperscript{a} & $\mathrm{rms}(r_\perp)$ for asymmetric cores, full residual for symmetric waves &
$\le 3\times10^{-2}$ \\
4 & virial identity, boundary-corrected & $< 10^{-2}$ \\
5 & evaluation-grid stability of the fixed network ($2\times$ in each direction) & $< 10\%$ \\
6\textsuperscript{a} & core stability on an extended domain ($L_{\rm ext}\simeq1.5L$) &
$|\Delta c|/c \le 2\times10^{-3}$. Field $\le 3\times10^{-2}$.
$r_\perp(L_{\rm ext}) \le 3\times10^{-2}$ \\
7 & mode-2 structure and speed range tested here (two extrema,
one node, $c\in(c_{\rm lw2},c_{\rm lw1})$, on-axis) & --- \\
8 & reproducibility from perturbed weights & field $\le 0.15$,
$|\Delta c|/c \le 3\times10^{-3}$ \\
9 & weakly nonlinear consistency at $A_{\rm small}$ (audit bound,
speed window, offset ordering) & per-profile \\
10 & residual decomposition, with $f_1$ reported and tested for
$O(a^2)$ scaling and domain dilution & factor-2 bands \\
\bottomrule
\end{tabular}

\vspace{4pt}
\begin{minipage}{\linewidth}
\footnotesize
\textsuperscript{a}The deep-water moderate-APE core fails the
$r_\perp$ bar of items 3, 6 and 10 and passes the others
(\S\ref{sec:mode2:scs}). Item 6
uses an extended domain of approximately $1.5L$, except that the
moderate benchmark core uses $L=9$ and $14$. The lower edge of the
item-9 speed band is half the equal-APE KdV speed offset. Unmeasured
items are identified as such.
\end{minipage}
\end{table}

\subsection{One start, three objectives}
\label{sec:verification:comparison}

The decisive comparison starts all three objectives from one network
fitted to the consistently normalized KdV wave \eqref{eq:kdvstart} at
$A_{\mathrm{ref}}$, with the same width-64 representation, $L=8$,
$512\times256$ quadrature, augmented-Lagrangian updates and speed
refresh. The fitted start carries the prescribed APE to
$2.26\times10^{-3}$, but it is not an admissible weakly nonlinear wave.
It overturns with $\max\eta_z=1.896$, it displaces fluid outside the
column with $\max(z-\eta)=1.126$, and its Rayleigh speed, $0.271$, is
subcritical and far below both the KdV formula speed, $0.551$, and the
reference speed, $0.433$. It is a finite-amplitude extrapolation of
weakly nonlinear theory, and a stringent test of selection.

\begin{figure}
  \centerline{\figpdf{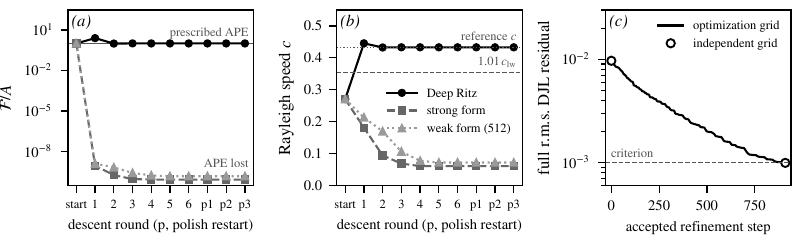}}
  \caption{One KdV-derived start, three objectives, at
  $A_{\mathrm{ref}}$. ($a$)~The APE ratio $\mathcal{F}/A$ and ($b$)~the
  Rayleigh speed at each saved state, which are the fitted start, the
  descent rounds 1--6 with multiplier and speed updates between rounds,
  and the refinement restarts p1--p3. Both residual objectives lose the
  APE within the first round and their Rayleigh speeds fall further
  below the supercriticality bound $1.01\,\clw$, whereas the Ritz
  objective overshoots in round 1 and settles on the reference speed.
  ($c$)~The subsequent constrained residual refinement of the Ritz
  endpoint (\S\ref{sec:mode1:polish}), showing the full r.m.s. residual
  along the 910 accepted steps on the optimization grid, the initial
  and final values on the independent $1024\times512$ grid (open
  circles) and the $10^{-3}$ criterion (dashed).}
  \label{fig:sharedstart}
\end{figure}

The outcomes separate cleanly (figure~\ref{fig:sharedstart}). The Ritz
objective recovers the depression wave, with relative APE, speed and
maximum field errors of $4.164\times10^{-7}$, $2.058\times10^{-6}$ and
$4.596\times10^{-4}$, respectively, and its multiplier converges to $5.333=1/c^2$,
the value required by \eqref{eq:el}. Its full residual,
$9.679\times10^{-3}$, nevertheless fails the accuracy criterion. However, the strong and weak (512-function)
objectives lose the APE in the first descent round and terminate at
$\mathcal{F}/A=8.64\times10^{-11}$ and $1.52\times10^{-10}$, respectively. Their
residuals, $2.324\times10^{-4}$ and $1.859\times10^{-4}$, pass the
accuracy criterion on their own, which is exactly why the residual criterion
is never applied on its own.

\subsection{Why the residual objectives lose the APE}
\label{sec:verification:mechanism}

Exact properties of the constrained problem help to interpret this
outcome, although they do not by themselves determine an optimizer's
path. First, the trivial field $\eta\equiv0$ is a stationary point of each augmented
Lagrangian \eqref{eq:al} used here, for every $(\lambda,\gamma)$,
because $\delta\mathcal{F}/\delta\eta=\eta N^2(z-\eta)$ by
\eqref{eq:dFdeta} and the first variations of $\mathcal{K}$,
$B_{\mathrm s}$ and $B_{\mathrm w}$ all vanish at $\eta=0$. Multiplier
and penalty updates change the objective value at a collapsed state but
create no first-order force at $\eta=0$. Nearby, the gradient is
proportional to the small remaining field amplitude, about $10^{-5}$ of
the wave's here. The record is consistent with this. As the multiplier
rises to about 970 and the penalty to its cap of $10^4$, the terminal
objective of both residual calculations grows to $32.75$, almost
entirely constraint terms, without any recovery of the APE.

Second, the amplitude direction through the start distinguishes the
objectives. Along the ray $s\eta_0$, with $P=\iint N^2(z-\eta)\eta^2$,
the derivative of $\mathcal{F}$ at $s=1$ is $P$, so the Ritz objective
has slope $2\mathcal{K}-\lambda_0P+\gamma_0(\mathcal{F}-A)P$. The
Rayleigh initialization gives $\lambda_0=2\mathcal{K}/P$ exactly, so
the start is stationary along this direction up to its small APE
error. For the residual objectives the slope is approximately
$2B-\lambda_0P$ when the dependence of $N^2$ on $\eta$ is neglected in
the derivative, about $74$ for the strong and $146$ for the weak
objective, so shrinking the field is a steep descent direction for
both. At the initial multipliers, $\lambda_0=13.6$ and $\gamma_0=200$,
abandoning the APE entirely costs $\lambda_0A+\tfrac12\gamma_0A^2\approx
0.49$, about $77$ and $150$ times below the start's $B_{\mathrm s}=37.2$
and $B_{\mathrm w}=73.5$. The wave itself, with zero residual, lies
lower still, so descent from this start can reach either state. For the Ritz objective
the trivial field lies above both the start ($\mathcal{K}=0.305$) and
the wave ($0.174$) at these multipliers. The failure of the weak form
is not a matter of visibility either, since
$B_{\mathrm w}/2B_{\mathrm s}=0.989$ and the 512-function space sees
$98.9\%$ of the start's squared residual. These statements concern the
local geometry and the energetic availability of states, not optimizer
trajectories. They are consistent with the observed collapse but do
not isolate selection from constraint enforcement, normalization and
speed updates. The comparison supports the expectation of
\S\ref{sec:formulations:strong} only for these protocols at this
single start. Two further tests with 512 global test functions show
that removing the positive initial multiplier does not restore
recovery. With $\lambda_0=0$, a reference-initialized calculation
retains the wave but fails the residual criterion, whereas a calculation
from the consistent start loses its APE in the first round, to
$1.5\times10^{-10}$ of the target.

\subsection{Reference-initialized calculations and test-space visibility}
\label{sec:verification:reference}

\begin{figure}
  \centerline{\figpdf{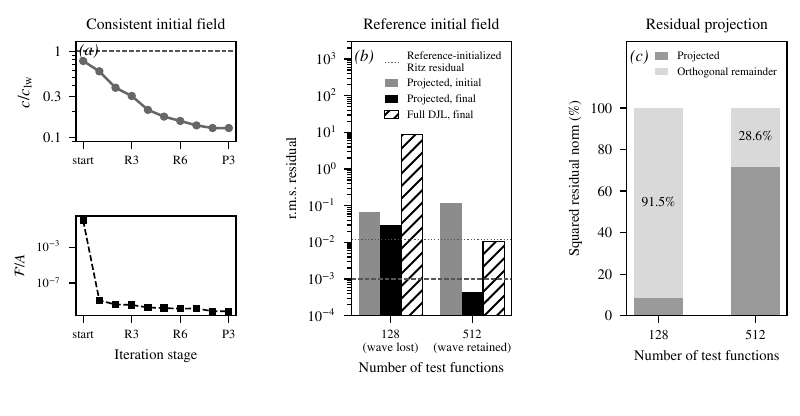}}
  \caption{Test-space visibility in the global weak formulation at
  $A_{\mathrm{ref}}$. ($a$)~Wave speed (upper panel) and APE ratio
  $\mathcal{F}/A$ (lower panel) during the 128-function calculation from
  the consistent start. In the first round the weak objective falls
  $128\times$ while the APE is lost. The dashed line marks the
  supercriticality bound.
  ($b$)~Reference-initialized calculations with 128 and 512 test
  functions. The wave is lost with 128 and retained with 512, for which
  the projected residual falls $270\times$ to $4.5\times10^{-4}$ while the
  full residual is $1.07\times10^{-2}$. The dotted line marks the
  reference-initialized Ritz residual and the dashed line the $10^{-3}$
  criterion. ($c$)~For a fixed reference-initialized Ritz field,
  $91.5\%$ and $28.6\%$ of the squared residual lie outside the 128- and
  512-function spaces.}
  \label{fig:v1split}
\end{figure}

Started from an approximation to the spectral reference, the residual
objectives can retain the wave, and their behaviour then measures the accuracy achieved rather than selection (figure~\ref{fig:v1split}$b$,$c$).
With 128 global test functions the weak form loses the solitary-wave
structure. In the first fixed-speed stage its objective falls
$16\times$ while the APE overshoots its target by $81\%$. With 512
functions the wave is retained and criteria 2--7 are met. The projected
r.m.s. residual falls $270\times$, to $4.5\times10^{-4}$, while the full
residual stays at $1.07\times10^{-2}$, because $99.8\%$ of the squared
residual lies outside the test space. Conversely, for a fixed
reference-initialized Ritz field the fractions of the squared residual
outside the 128- and 512-function spaces are $91.5\%$ and $28.6\%$, so
enlarging the space exposes more of a given residual without
guaranteeing a smaller residual after optimization. The direct
strong-form minimization reaches $8.7\times10^{-4}$, against
$1.21\times10^{-2}$ for the Ritz objective, in the order anticipated.
These are measured endpoints, not bounds, and the quantitative bands
used to judge them are not fixed in advance. An
element-based test space with 512 local functions collapses the APE to
$5.3\times10^{-7}$ of its target even from the reference
approximation, whose weak objective, $0.92$, is nearly twice that of
the 512 global functions, $0.48$. Resolving the residual better makes
the trivial state more available, not less, because the global minimum
of every residual objective is unchanged.

From the consistent start the 128-function weak form fails in the same
way as the 512-function form. In its first round the weak objective
falls $128\times$ while the APE falls to $1.1\times10^{-9}$ of its target
(figure~\ref{fig:v1split}$a$), so decreasing a projected residual
here means abandoning the wave rather than approaching it. A second
Ritz initialization from the same start recovers the wave, with speed
and field errors of $4.1\times10^{-6}$ and $7.0\times10^{-4}$ and a
residual of $1.16\times10^{-2}$.

\begin{table}
\centering
\caption{Properties, expectations and measurements for the three
objectives at $A_{\mathrm{ref}}$. The expectations are fixed before the measurements and apply to
the shared start, the consistently normalized but physically inadmissible
KdV-derived field of \S\ref{sec:verification:comparison}. Here $r_{\rm rms}$
is the full r.m.s. DJL residual on the independent grid, order is the
highest derivative of the trial field in the objective, and selection
denotes an energetic ranking of exact roots, not a convergence
guarantee. The strong-form reference-initialized value uses the exploratory collocation procedure of Appendix~\ref{app:numerics}, and the Ritz value uses width 128.}
\label{tab:taxonomy}
\small
\setlength{\tabcolsep}{4pt}
\begin{tabular}{@{}P{1.75cm}P{0.95cm}P{1.6cm}P{1.3cm}P{2.4cm}P{1.5cm}P{2.1cm}@{}}
\toprule
 & & & \multicolumn{2}{c}{Shared start} & \multicolumn{2}{c}{Reference-initialized $r_{\rm rms}$} \\
\cmidrule(lr){4-5}\cmidrule(l){6-7}
Objective & Order & Selection & Expected & Observed & Expected & Observed \\
\midrule
Strong-form residual & 2nd & none & fails &
APE lost, $\mathcal{F}/A=8.64\times10^{-11}$, $r_{\rm rms}=2.324\times10^{-4}$ &
lowest & $8.7\times10^{-4}$ \\
Weak-form residual (512) & 1st & none & fails &
APE lost, $\mathcal{F}/A=1.52\times10^{-10}$, $r_{\rm rms}=1.859\times10^{-4}$ &
above strong, improving with test space &
wave lost with 128, $1.07\times10^{-2}$ with 512 \\
Constrained Deep Ritz & 1st & Turkington & recovers &
wave recovered, $r_{\rm rms}=9.679\times10^{-3}$ &
above strong & $1.21\times10^{-2}$ \\
\bottomrule
\end{tabular}
\end{table}

\subsection{Access from unstructured fields and across amplitude}
\label{sec:mode1:reachability}

\begin{figure}
  \centerline{\figpdf{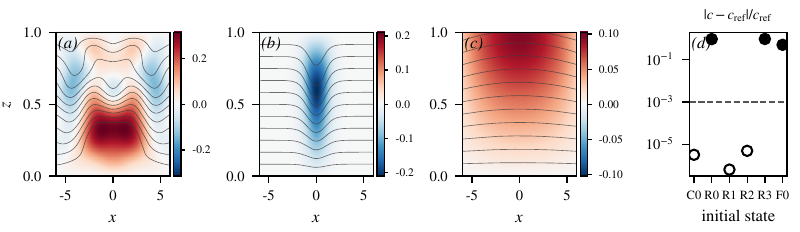}}
  \caption{Recovery from random initial fields at $A_{\mathrm{ref}}$.
  Panels ($a$)--($c$) show the displacement $\eta(x,z)$ in colour with
  displaced isopycnals. ($a$)~A random field carrying the prescribed APE
  and satisfying the boundary conditions and horizontal symmetry of the
  representation. ($b$)~The wave it descends to, whose speed error is
  $6\times10^{-7}$. ($c$)~The near-lid state reached from another
  random start. Its $\mathcal{K}$ on the optimization grid undercuts the
  wave's, and the refined-grid check rejects it. ($d$)~Speed error of
  every start against the $10^{-3}$ threshold. Open circles reach the
  wave and filled circles do not. C0 is the consistent start,
  R0--R3 are the random starts and F0 is the consistent start with
  its sign reversed.}
  \label{fig:genericstarts}
\end{figure}

The Ritz objective does not depend on the start resembling a wave.
Four smooth random fields, each a sum of twenty-four low-order modes
with normally distributed coefficients and no imposed wave structure,
are rescaled to carry the target APE and submitted to the same
protocol, with the physical checks disabled because they refer to the
initial state. Two of them, one initially of elevation polarity,
descend onto the wave, with speed errors of $6.3\times10^{-7}$ and
$4.8\times10^{-6}$ beside maximum field errors of $2.6\times10^{-4}$
and $7.1\times10^{-4}$, and with residuals $0.69$ and $0.99$ times the
reference-initialized value (figure~\ref{fig:genericstarts}). The other
two end on the same object, a
broad elevation pressed against the lid with its extremum in the top
grid cell. It is not a rival wave. On the optimization grid its
$\mathcal{K}$ is $0.034$--$0.038$ against the wave's $0.174$ at the same
APE, but on the independent grid its Rayleigh speed falls from
$0.88$--$0.95$ to $0.014$--$0.020$ and its residual lies five orders
of magnitude above the reference-initialized value. Across the
$512\times256$, $1024\times512$ and $2048\times1024$ grids its speed
moves by factors of $1.5$ to $4.6$, where genuine states agree to six
digits, so a single refined-grid comparison identifies this
finite-quadrature artefact without consulting any reference. The
consistent start with its sign reversed, an elevation that overturns
with $\max|\eta_z|\approx10$, does not reach the wave either. It stalls on
a depression at mid-depth whose $\mathcal{K}$ is $4.7$ times the wave's,
so it is a failed descent rather than a competing state. A sample of four is too small
to map the wave's basin of attraction or to estimate a success rate,
and classical random-start computations are not compared.

\begin{figure}
  \centerline{\figpdf{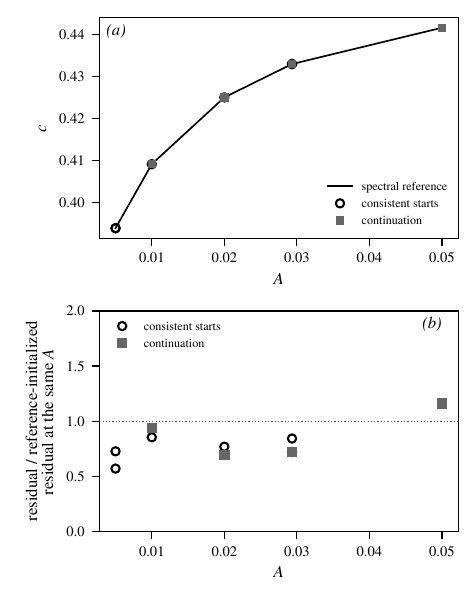}}
  \caption{The mode-1 wave family from the consistent start and by
  continuation. ($a$)~$c(A)$ against the spectral reference.
  ($b$)~Residual divided by the reference-initialized residual at the
  same amplitude and width, with ratios from $0.57$ to $1.16$.}
  \label{fig:dr3branch}
\end{figure}

The wave family is computed in two ways. First, independent calculations
start from the KdV-derived field at prescribed APE values
$A=0.005$, $0.010$, $0.020$ and $A_{\mathrm{ref}}$, with two network
initializations at the smallest and reference values. Second, continuation
follows the family from $A=0.005$ to $0.050$, using the converged wave and
APE multiplier at each step to initialize the calculation at the next,
larger APE. This approach also reaches $A=0.050$, beyond the APE range of
the KdV initialization used here (\S\ref{sec:formulations:kdv}). All six direct and four
continuation calculations satisfy the wave-structure criteria 2--7 without
violating the physical checks, although their full DJL residuals remain
above the accuracy threshold. Their residuals are comparable to those
obtained by initializing the Ritz calculation from a spectral reference
wave, the ratio being $0.57$--$1.16$ at the same APE (figure~\ref{fig:dr3branch}). The relative speed
error decreases from $1.3\times10^{-5}$ at $A=0.005$ to
$3.4\times10^{-6}$ at $A=0.050$, with corresponding field errors of
$1.3\times10^{-3}$ and $5.6\times10^{-4}$. This trend is consistent with
the energy landscape in \S\ref{sec:djl:landscape}, in which the energetic penalty for broadening at
fixed APE is weakest for small waves, making their width less strongly
selected by the objective. The few calculations available, however, do
not establish that this mechanism causes the decrease in speed error.
Where both approaches are tested, continuation agrees with the direct
calculations to within $|\Delta c|\leq1.3\times10^{-6}$, at 29\% lower
computational cost.

The scope of these results is stated here once. They establish
selection for one stratification under one fixed protocol, from the
consistent start at four amplitudes and by continuation to a fifth.
The random starts show that the wave is reached from fields with no
wave structure, without measuring the size of its basin. Profiles
indistinguishable at KdV order can have different fully nonlinear
fates, so the dependence on stratification remains open.

\subsection{Competing states}
\label{sec:mode1:uniqueness}

A deflated search, which discourages convergence back to the known wave,
is performed from five prescribed initial fields within the $x$-symmetric representation (Appendix~\ref{app:uniqueness}). It produces no additional accepted wave.
The two candidates that remain stable under grid refinement are a broadened
single-core field, denoted $X_1$, and a compressed single-core field,
denoted $X_2$, with half-widths respectively 1.49 and 0.78 times that of
the known wave. At the same prescribed APE, their objective values satisfy
\[
  \mathcal{K}(\text{wave})=0.17372<\mathcal{K}(X_1)=0.17567
  <\mathcal{K}(X_2)=0.17589 .
\]
Both therefore cost only about 1\% more $\mathcal{K}$ than the wave,
despite their appreciably different widths, and their objective values
agree within 1\% with those of the dilation family in \S\ref{sec:djl:landscape} at the
same widths. Neither candidate satisfies the virial identity. Their
discrepancies are $6$--$9\times10^{-2}$, roughly $200\times$ the largest
value recorded for the solitary waves. These fields thus illustrate again
the weak preference for the wave's width in the energy landscape.
Every non-degenerate
terminal carries more $\mathcal{K}$ at equal APE than the wave, a
direct witness of the selection principle at work, though not a proof
of uniqueness.

\section{Accuracy, from a recovered wave to a resolved equation}
\label{sec:mode1}

\subsection{Why the Ritz residual stalls}
\label{sec:mode1:floor}

The Ritz waves recover structure, speed and APE but leave residuals
near $10^{-2}$, $10\times$ above the criterion, and the residual proves insensitive to the obvious remedies of further optimization, a wider network, a different far-field factor in the ansatz and a change of quadrature resolution. At $A_{\mathrm{ref}}$
a further full optimization pass of the reference-initialized width-64
field leaves its residual at $1.2258\times10^{-2}$ to five significant
figures, quadrupling the parameter count (width 128) buys only $1.5\%$,
to $1.208\times10^{-2}$, a different far-field factor does not help, and
coarsening the quadrature makes it worse. At the other four amplitudes,
the single reference-initialized calculations at equal iteration count
give $8.94\times10^{-3}$ to $1.389\times10^{-2}$, a non-monotone spread
of $1.55\times$ that must be read against the $1.30$--$1.42\times$
scatter between network initializations at fixed amplitude, so no
systematic amplitude dependence is resolved.

\begin{figure}
  \centerline{\figpdf{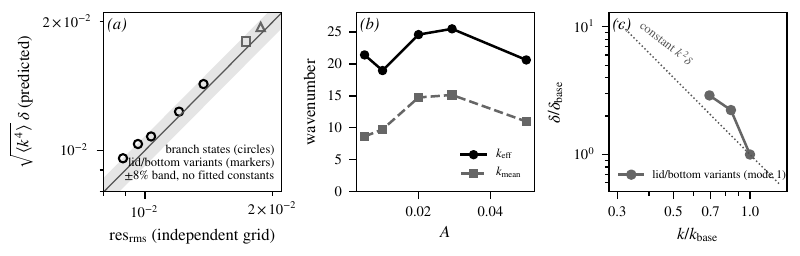}}
  \caption{The residual stall in $H^2$ form. ($a$)~The estimate
  $\sqrt{\langle k^4\rangle}\,\delta$ against the independently measured
  residual at seven recorded states, with no fitted constant. ($b$)~The
  fourth-moment wavenumber $k_{\mathrm{eff}}$ and the mean wavenumber
  $k_{\mathrm{mean}}$ against amplitude. Their ratio is largest at the
  smallest amplitude. ($c$)~Changes in error wavenumber
  and amplitude under adaptation at the rigid lid and bottom, with $k$
  the mean wavenumber $k_{\mathrm{mean}}$ and $\delta$ measured. The panels describe the residual
  that the measured errors produce, not the cause of their persistence
  under further iteration.}
  \label{fig:floorlaw}
\end{figure}

The mechanism of the stall is measured, and in a form that survives a
test on states it is not built from (figure~\ref{fig:floorlaw}). With
$d=\eta_\theta-\eta^{\mathrm{ref}}$ and $S(\eta,c)=N^2(z-\eta)\eta/c^2$,
subtracting \eqref{eq:djl} for the two fields gives exactly
\[
  r_\theta=\nabla^2 d+S(\eta_\theta,c_\theta)-S(\eta^{\mathrm{ref}},c_{\mathrm{ref}})
  +r_{\mathrm{ref}} .
\]
Because the reference residual and the source difference are
small, the residual of an optimized state is, to a few per cent, the
Laplacian of its error field. A boundary-compatible orthonormal
expansion of $d$ then gives
\begin{equation}
  \mathrm{res}_{\mathrm{rms}}\approx k_{\mathrm{eff}}^2\,\delta,\qquad
  k_{\mathrm{eff}}=\langle k^4\rangle^{1/4},\qquad
  \delta=\mathrm{rms}(d),
  \label{eq:floorlaw}
\end{equation}
with the fourth moment weighted by the squared error coefficients. The
residual is the $H^2$ seminorm of the error, whereas the Ritz objective
controls its $H^1$ seminorm, and the gap between what is controlled and
what is scored is one factor of an effective wavenumber. Evaluated on
the recorded error fields of five reference-initialized waves and two
boundary-adapted variants, \eqref{eq:floorlaw} predicts the
independently measured residual to $2.4$--$8.0\%$ with no fitted
constant, and to $0.6$--$3.4\%$ in an orthonormal re-evaluation. The
effective wavenumber spans $18.1$--$25.5$, but its ratio to the mean
wavenumber varies from $1.7$ to $2.5$ (panel $b$), so that a scalar
compression of the law built on $k_{\mathrm{mean}}^2\delta$ misses the
smallest amplitude by a factor of $2.1$, because the fourth moment is dominated
by a spectral tail which the mean does not see.

If the error spectrum were unchanged, \eqref{eq:floorlaw} would turn
the criterion into a requirement on the size of the error, since at the
measured $k_{\mathrm{eff}}$ a residual of $10^{-3}$ needs an r.m.s.
displacement error of $1.5$--$3.1\times10^{-6}$ of the depth, an order
of magnitude below that of the Ritz fields. The refinement of
\S\ref{sec:mode1:polish} shows that the criterion is instead met by
changing the spectral content of the error. A reading consistent with
that outcome, though not tested by a spectral measurement of the
refined field, is that high-wavenumber error contributes little to
$\mathcal{K}$, which weights the error spectrum by $k^2$, whereas the
residual weights it by $k^4$. The boundary is not the obstacle. The
factor $z(1-z)$ in \eqref{eq:ansatz} does not enforce $\eta_{zz}=0$ at
lid and bottom, and the nodes there, $0.8\%$ of the grid, carry
$2.6$--$18.8\%$ of the squared residual. A representation that enforces
zero boundary curvature drives the wall residual to roundoff but makes
the total $1.41\times$ worse, from $1.2258\times10^{-2}$ to
$1.7310\times10^{-2}$, because the spectrum moves favourably and the
error amplitude grows faster (panel $c$). Within the vertical-factor
family tested at fixed width, $k$ and $\delta$ are coupled through the
ansatz and no tested factor moves their product favourably.

\subsection{Constrained residual refinement}
\label{sec:mode1:polish}

If the deficit is one of spectral weighting rather than
representation, a residual objective applied to the recovered wave
should remove it, and it does. The Ritz wave recovered from the consistent start in
\S\ref{sec:verification:comparison} is refined with the full DJL residual as objective, the same squared
residual that the strong form minimizes in
\S\ref{sec:verification:comparison}, but with the APE held exactly and
the speed differentiated rather than frozen, retaining the width-64 representation, $L=8$ and
the $512\times256$ quadrature. The APE is enforced by a scalar
normalization and the speed is recomputed from the Rayleigh quotient at
every evaluation, with derivatives taken through both
(Appendix~\ref{app:numerics}), so that collapse is excluded by
construction. After 910 accepted steps and 1,869 objective evaluations, the residual falls from $9.679\times10^{-3}$ to
$9.98581\times10^{-4}$ on the $1024\times512$ grid and
$9.98619\times10^{-4}$ on the $2048\times1024$ grid
(figure~\ref{fig:sharedstart}$c$), and the state meets every
acceptance criterion, with relative APE, speed and maximum field errors
of $1.28\times10^{-10}$, $2.33\times10^{-6}$ and $4.84\times10^{-4}$.
A second refinement under the same protocol, from an independently
fitted start, passes the same joint criteria, with residuals of
$9.996\times10^{-4}$ on both grids after 880 accepted steps.

The changes in the other diagnostics are as revealing as the residual.
The field moves by only $0.0523\%$ in maximum norm, yet the residual
falls $9.7\times$. Meanwhile the relative $L^2$ field error against the
spectral reference rises $1.9\times$, from $4.62\times10^{-4}$ to
$8.71\times10^{-4}$, the speed error rises from $2.06\times10^{-6}$ to
$2.33\times10^{-6}$, and the virial discrepancy rises $5.7\times$, from
$1.62\times10^{-5}$ to $9.28\times10^{-5}$. A residual that falls while
$\delta$ grows means that the improvement comes from the spectral
content of the error rather than its size, consistent with
\eqref{eq:floorlaw}. Small-scale error that the residual weights
heavily is removed while the field drifts at the large scales that
dominate the comparison with the reference. The Ritz residual is
therefore not an accuracy limit of this representation. The two
objectives play complementary roles, the energy selecting the wave and
the residual resolving it, in the manner of the two-stage strategy of
\citet{song2024}, except that here the first stage is itself a
constrained minimization that also reaches the wave from random fields
without imposed wave structure. The result concerns one configuration,
establishes neither constrained stationarity nor direct residual
recovery, and uses a constraint and speed treatment different from
those of the direct comparison. In the second refinement the secondary
diagnostics move differently, the $L^2$ error rising by only
$2\%$ while the speed and maximum field errors improve, so the
direction of these small changes is not a reproducible feature.

\section{Mode-2 internal solitary waves}\label{sec:mode2}

\subsection{The physical problem}
\label{sec:mode2:problem}

A mode-2 internal solitary wave displaces isopycnals in opposite
directions above and below the pycnocline, characterized by a double-lobed structure
with one interior node. It usually travels between
the mode-2 and mode-1 linear long-wave speeds, $\clw[2]<c<\clw[1]$.
Such waves have been observed in the laboratory since
\citet{davis1967}, on the New Jersey shelf \citep{shroyer2010} and on
the northern slope of the South China Sea \citep{yang2009}.
Mathematically they are harder objects than their mode-1 relatives. A wave inside the mode-1
speed window is \emph{embedded} in the mode-1 radiation spectrum, so an
exactly steady, exactly localized mode-2 solitary wave does not
generically exist. The generic object is a core with a small oscillatory
mode-1 tail \citep{akylas1992}, exactly localized embedded waves arise
in three-layer flows only at isolated parameter values
\citep{doak2022}, and offset pycnoclines produce asymmetric mode-2
waves with mode-1 tails in the laboratory \citep{carr2015} and in
direct simulation \citep{olsthoorn2013}. The physically relevant
questions are therefore quantitative. How small is the tail, and over
what distance does the radiation drain the wave?

For a variational solver the embedding also redoubles the selection
problem anticipated in \S\ref{sec:djl:classical}. At equal APE, the mode-1 wave has a lower $\mathcal{K}$ by a factor
of $9$--$11$ on the symmetric profile below and at least $7$ on the
asymmetric one, so unrestricted minimization favours mode 1. These are ratios of
$\mathcal{K}$ against trial-family energies, not of $E_K$, and they
imply no convergence guarantee from arbitrary fields, but they make
plain that computing mode 2 variationally requires excluding the
mode-1 branch from the search space itself.

\subsection{Structural exclusion of mode 1}
\label{sec:mode2:formulations}

On a stratification symmetric about mid-depth the exclusion is
structural. Antisymmetrizing the network output,
\begin{equation}
  \eta_\theta=z(1-z)(1-\hat x)\bigl[\mathcal{N}_\theta(\hat x,z)
  -\mathcal{N}_\theta(\hat x,1-z)\bigr],
  \label{eq:antisym}
\end{equation}
makes every representable field antisymmetric about $z=\tfrac12$. 
Mode-1 content is outside the search space throughout optimization,
and the stratification symmetry ensures that the mode-1 forcing
vanishes identically. The target is therefore an exactly localized wave
of the unforced DJL equation \eqref{eq:djl}, which the computed fields approximate to
the reported accuracy. The symmetric configuration provides a controlled
diagnostic of mode-2 computation without mode-1 forcing. For asymmetric ocean
stratifications, mode 1 is excluded by a hard projection incorporated
into the neural representation and differentiated during optimization,
\begin{equation}
  \eta_\theta=\eta_{\mathrm{raw}}-\phi_1(z)\,
  \frac{\langle\eta_{\mathrm{raw}}(x,\cdot),N^2\phi_1\rangle}
       {\langle\phi_1,N^2\phi_1\rangle},
  \label{eq:projansatz}
\end{equation}
with $\eta_{\mathrm{raw}}$ the ansatz \eqref{eq:ansatz} and $\phi_1$
the long-wave mode-1 eigenfunction of \eqref{eq:lw}. The $N^2$-weighted
inner product is the one under which the linear modes are mutually
orthogonal, so the constraint is inactive at leading order. With
$w=N^2\phi_1/\|N^2\phi_1\|$, minimizing $\mathcal{K}$ at $\mathcal{F}=A$
under this restriction gives the forced Euler--Lagrange equation
\begin{equation}
  r\equiv\nabla^2\eta+\frac{N^2(z-\eta)}{c^2}\eta=f_1(x)\,w(z) ,
  \label{eq:el2}
\end{equation}
in which $f_1$ is the multiplier force of the orthogonality constraint.
The projected object is a constrained \emph{quasi-steady core}, held
steady by that force rather than an exact solution of the unforced DJL equation. Its residual provides the source used to estimate mode-1
radiation. This dictates the adaptation of the
acceptance criteria (table~\ref{tab:certificates}). The residual is
decomposed into $f_1w$ and the orthogonal remainder $r_\perp=r-f_1w$.
Only $r_\perp$ is subject
to the residual threshold, while $f_1$ is reported separately and tested
for two expected consistency properties, approximately quadratic scaling with the core displacement
amplitude $a$, $f_1=O(a^2)$, and stability of the overall forcing strength
under extension of the horizontal computational domain.

\subsection{Radiation from a prescribed core}
\label{sec:mode2:radiation}

The tail is a derived quantity, estimated by treating the computed core
as a prescribed source of linear mode-1 radiation. At the speed of the
core the resonant mode satisfies
\begin{equation}
  \Bigl[-\partial_{zz}-\frac{N^2(z)}{c^2}\Bigr]\psi_1=-k_1^2\psi_1,\qquad
  \psi_1(0)=\psi_1(1)=0,\qquad\int_0^1\psi_1^2\,\mathrm{d}z=1 ,
  \label{eq:resonantmode}
\end{equation}
and a cancelling response $a(x)\psi_1(z)$ obeys $a''+k_1^2a=f_{\mathrm{osc}}$
with $f_{\mathrm{osc}}=-\int_0^1r\psi_1\,\mathrm{d}z
=-f_1\langle w,\psi_1\rangle-\langle r_\perp,\psi_1\rangle$, so the
full residual, not only the multiplier force, drives the tail. The core is approximated as horizontally symmetric, with the feedback
of the radiation on its shape neglected, so the forcing is even,
$f_{\mathrm{osc}}(x)=f_{\mathrm{osc}}(-x)$, while the radiated correction
itself may be asymmetric. Since the resonant mode-1 waves have group
velocity $c_g<c$, their energy propagates behind the core, and imposing
no incoming radiation ahead of it gives a trailing tail with modal and
peak amplitudes
\begin{equation}
  a_{\mathrm{out}}=\frac{2}{k_1}\Bigl|\int_0^Lf_{\mathrm{osc}}(x)\cos(k_1x)
  \,\mathrm{d}x\Bigr|,\qquad A_{\mathrm{tail}}=a_{\mathrm{out}}\max|\psi_1| .
  \label{eq:tailamplitude}
\end{equation}
With $J=\int_0^1N^2\psi_1^2\,\mathrm{d}z$,
the wavelength-averaged tail energy, the radiated power and the
fractional energy loss over a distance $D$ are
\begin{equation}
  \overline E_{\mathrm{tail}}=\tfrac12a_{\mathrm{out}}^2J,\qquad
  P_{\mathrm{rad}}=(c-c_g)\overline E_{\mathrm{tail}},\qquad
  \varepsilon(D)\simeq\frac{P_{\mathrm{rad}}D}{c\,E_{\mathrm{core}}},
  \label{eq:radiativeloss}
\end{equation}
with $E_{\mathrm{core}}=c^2\mathcal{K}+\mathcal{F}$ and group velocity
$c_g=c(1-k_1^2c^2/J)<c$, so that the tail is phase-locked to the core
while its energy falls behind. Every tail estimate is reported against
the coherent noise level of the diagnostic, built from the far-window
behaviour of $f_{\mathrm{osc}}$, and is called resolved only when the
readings on two independently computed domains both exceed that level
and agree within a factor of two. Otherwise it is an unresolved
diagnostic, not a bound. An aggregate sensitivity analysis yields
conditional loss estimates (Appendix~\ref{app:numerics}) that assume
unobserved errors in the resonant source not exceeding the assessed
sensitivity. The estimate neglects feedback of the radiation on the
core, it is not validated against time-dependent simulation at these
amplitudes, and agreement between domains does not bound errors shared
by both. These caveats travel with every radiation number below.

\subsection{Cores on four stratifications}
\label{sec:mode2:rungs}\label{sec:mode2:scs}

\begin{figure}
  \centerline{\figpdf{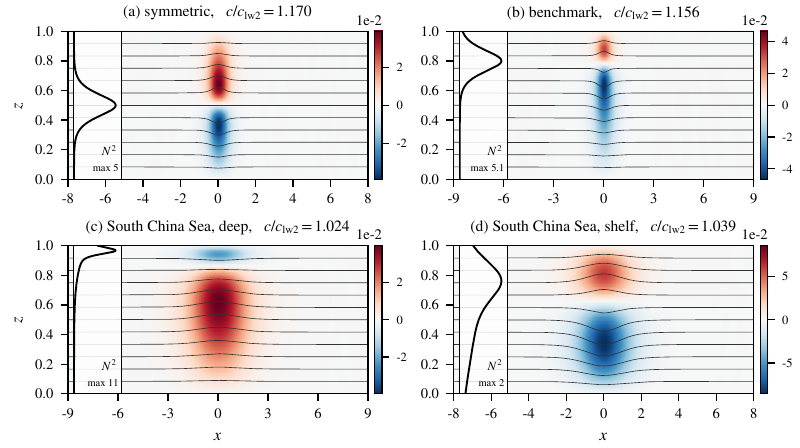}}
  \caption{Mode-2 displacement fields $\eta(x,z)$ for the
  $A_{\mathrm{mod}}$ wave or core on the four stratifications,
  mirrored about $x=0$, with displaced isopycnals at equally spaced
  upstream heights. The terminal $c/\clw[2]$ is given above each panel,
  and the inset shows the squared buoyancy frequency $N^2(z)$ on a
  linear scale from zero to the stated maximum, at the same $z$ as the
  field. ($a$)~Symmetric,
  ($b$)~benchmark, ($c$)~South China Sea deep and ($d$)~South China Sea
  shelf profiles. The symmetric, benchmark and shelf profiles give
  convex waves and the deep profile a concave core. The symmetric case
  approximates a localized DJL wave, whereas the asymmetric cases are
  projected quasi-steady cores.}
  \label{fig:mode2gallery}
\end{figure}

The study computes two waves on each of three stratifications and one
on a fourth
(table~\ref{tab:mode2}, figure~\ref{fig:mode2gallery}). Each
stratification carries two prescribed APE values, $A_{\rm small}$ and
$A_{\rm mod}$ (table~\ref{tab:mode2}), at which the weakly nonlinear KdV
wave has a maximum isopycnal slope $|\partial\eta/\partial z|$ of about
$0.2$ and $0.55$--$0.6$, respectively. The computed waves reach similar
slopes, $0.18$--$0.20$ and $0.50$--$0.57$, so the two values sample a
weakly and a moderately steep wave on each profile. Every core starts
from the consistent KdV width with a mode-2 vertical structure.

\emph{Symmetric diagnostic configuration.} For a pycnocline symmetric about mid-depth, the antisymmetric
representation \eqref{eq:antisym} yields two verified approximations
to localized mode-2 waves. Both are embedded in the mode-1 radiation
spectrum, with $c/\clw[1]=0.31$--$0.35$. At the larger amplitude,
extending the horizontal domain by a factor of 1.5 changes the speed
by only $|\Delta c|/c=3.3\times10^{-4}$, consistent with localization. The weakly nonlinear description is
close across all seven mode-2 endpoints, with half-widths $0.92$--$1.12\times$
 those of the equal-APE KdV waves and speeds within $-0.50\%$ to
$+1.54\%$ of theirs. Mid-depth symmetry does not make the two polarities equivalent at fixed
APE. The convex
wave, whose isopycnals above the interior node rise and those below
descend, and the concave wave, its reverse, differ in $\mathcal{K}$ by
$26\%$ here in an equal-APE trial-field probe. For all four stratification profiles, the trial field with the polarity
predicted by the sign of $r_{10,2}$, the mode-2 analogue of the KdV
coefficient $r_{10}$, has a smaller $\mathcal{K}$ than the field of
opposite polarity at the same APE. Because the initialization sets
the polarity and the mode-2 checks record rather than reject reversals,
this agreement is not an independent optimization of both polarities.

\emph{Asymmetric benchmark profile.} On the benchmark profile of
\S\ref{sec:djl:landscape}, whose pycnocline is offset towards the
surface, the projected formulation produces two separate quasi-steady mode-2 wave
cores, one at each prescribed APE, $A_{\mathrm{small}}$ and
$A_{\mathrm{mod}}$. Figure~\ref{fig:mode2gallery}$b$ shows only the
$A_{\mathrm{mod}}$ core, including both its upper and lower displacement
lobes. For
this steeper core, the full residual exceeds the orthogonal-remainder
band entirely through $f_1$, which behaves as a physical forcing must.
It scales as the square of the amplitude, with a measured exponent
ratio $S=1.02$ inside a factor-2 band, and its total is invariant under
domain extension, $D=1.001$, while $r_\perp=1.8\times10^{-2}$ sits
inside its band. The excess is the physics of the embedded wave, not a
defect of the representation. This core is also one of the two for
which the tail is resolved. Its peak outgoing displacement is $2.9\%$
of the core amplitude on the base domain and $2.7\%$ on the extended
one, and it loses $3.2\%$ and $2.7\%$ of its energy over 100
half-widths, with a conditional estimate of up to $4.9\%$. The $A_{\mathrm{small}}$ core on this profile has an unresolved tail, whose conditional loss over the same distance is below $10^{-3}\%$.

\emph{The observation-fitted profile, deep basin.} The target
stratification is fitted to the annual-mean World Ocean Atlas
temperature and salinity at $20.5^\circ$N, $120.5^\circ$E in the South
China Sea, from which the potential density and the squared buoyancy
frequency are computed and interpolated to 301 depth levels down to
$H=3000$~m. A two-component $\tanh$ fit reproduces the mode-1 and
mode-2 long-wave speeds to $9.7\times10^{-4}$ and $3.8\times10^{-5}$,
the scaled density to $3.5\times10^{-3}$ and the $N^2$ profile with a
relative $L^2$ error of $15\%$. Everything
computed on this stratification is a property of that fitted annual
mean. Mode-2 activity in the region is strongly seasonal, the
observations quoted below finding mode-2 waves predominantly in winter,
and a seasonal refit would move the linear windows, the dimensional
speeds and, in principle, the sign of $r_{10,2}$, a sensitivity untested
here. On this profile the polarity flips relative to the two previous
stratifications, $r_{10,2}=-0.769$ selecting the concave side. Two
separate quasi-steady cores are computed at $A_{\mathrm{small}}$ and
$A_{\mathrm{mod}}$. The $A_{\mathrm{small}}$ core travels at
$1.4388$~m\,s$^{-1}$ with a displacement of $41.8$~m and a half-width
of $7.15$~km, and it passes every measured item. The
$A_{\mathrm{mod}}$ core, shown in figure~\ref{fig:mode2gallery}$c$
with both its displacement lobes, is the centrepiece of the study and
repeats the mode-1 division of labour between selection and resolution.
Its Ritz core displaces isopycnals by $119.5$~m, travels at
$1.4601$~m\,s$^{-1}$ inside the linear window
$(1.4265,2.7020)$~m\,s$^{-1}$, and has a half-width of $4.09$~km,
$2\%$ wider than the equal-APE KdV wave, its node at $414$~m and its
extremum near $1150$~m (table~\ref{tab:scs}). Its virial discrepancy
is $2.4\times10^{-4}$, but its orthogonal remainder,
$6.4\times10^{-2}$, fails the $3\times10^{-2}$ criterion, and
$84\%$ of $r_\perp^2$ lies in the top $5\%$ of the column, where the
thin surface pycnocline is resolved least well. The residual objective
can remove the defect, but not without moving the wave. A constrained
residual refinement of the kind used in \S\ref{sec:mode1:polish}, with
$r_\perp$ as objective and stopped as soon as the criterion is met,
reaches $r_\perp=2.9\times10^{-2}$ with $\mathcal{K}$ only $0.1\%$
higher, but the core is then $19\%$ wider and $8\%$ lower, and its
virial discrepancy has risen ninefold. Continued to
$1.7\times10^{-2}$, the refinement widens the core by $36\%$. The
energy is nearly flat along the width direction that the residual
objective follows, the weak width preference of
\S\ref{sec:djl:landscape} seen again in mode 2, so the
$A_{\mathrm{mod}}$ Ritz core, the state the selection principle picks,
is the reported wave, and its residual is its one acknowledged defect.
Its tail is unresolved, with a conditional loss over 100 half-widths
below $10^{-5}$.

\emph{The shelf configuration.} Truncating the same profile at
$350$~m, the depth of the isobath at which \citet{yang2009} recorded
mode-2 internal solitary waves. The pycnocline, where $N^2$ peaks at $70$~m depth, now lies
one-fifth of the way down the $350$~m shelf column, compared with
approximately one-fortieth of the $3000$~m deep-basin column.
At the seabed, $N^2$ remains about one-sixth of its peak value,
so stratification persists through the full water column
(figure~\ref{fig:mode2gallery}$d$). The computed wave
displaces isopycnals by $29.9$~m and travels at $0.6225$~m\,s$^{-1}$,
$4\%$ above the mode-2 long-wave speed and inside the window
$(0.5991,1.2780)$~m\,s$^{-1}$. Its half-width is $387$~m, and its two
lobes, a $29.9$~m depression at $230$~m depth and a $21.1$~m elevation
at $68$~m, straddle a node at $132$~m. The polarity flips back with the
truncation, $r_{10,2}=+0.735$ against the deep column's $-0.769$, and, in the equal-APE trial-field comparison, the predicted convex
polarity yields a $13.1\%$ lower $\mathcal{K}$ than the opposite polarity. Which way a mode-2 wave of given APE deforms the pycnocline
therefore depends on the entire stratification, not only on the
pycnocline itself. The core passes all ten items of
table~\ref{tab:certificates} with an orthogonal remainder of
$1.5\times10^{-2}$, and a perturbed restart returns to the same wave to
$2.7\times10^{-4}$ in field. It ends at the prescribed iteration count
without meeting the stricter gradient tolerance of the stationarity
test. Its tail is the second resolved estimate, $9.1$~cm or $0.30\%$ of
the core at a resonant wavelength of $355$~m, with an energy loss over
100 half-widths of $0.027\%$, or $0.030\%$ under the sensitivity-based loss estimate.

\begin{table}
\centering
\caption{The seven mode-2 endpoints. $A$ is the prescribed APE. The
items give the number of items passed over items measured
(table~\ref{tab:certificates}). $k_1W$ is the resonant wavenumber times
the core half-width. Tail ratios are peak outgoing displacement over
core amplitude on base/extended domains, and losses are fractions of
core energy over 100 half-widths. Values in parentheses are conditional
estimates under the aggregate-sensitivity assumption, not bounds.
Radiation is assessed separately from acceptance of the core, and the
antisymmetric waves have none, since $f_{\rm osc}\equiv0$.}
\label{tab:mode2}
\small
\setlength{\tabcolsep}{4pt}
\begin{tabular}{@{}lllllll@{}}
\toprule
Profile & Wave & $A$ & Items & $k_1W$ & $A_{\rm tail}/a_{\rm core}$ & Loss \\
\midrule
symmetric & $A_{\rm small}$ & $4.897\times10^{-5}$ & 9/9 & --- & none & --- \\
symmetric & $A_{\rm mod}$ & $2.795\times10^{-4}$ & 8/8 & --- & none & --- \\
benchmark & $A_{\rm small}$ & $4.360\times10^{-5}$ & 10/10 & 9.3 & unresolved & ($0.0005\%$) \\
benchmark & $A_{\rm mod}$ & $2.449\times10^{-4}$ & 9/9 & 4.6 & $2.9\%/2.7\%$ & $3.2\%/2.7\%$ ($4.9\%$) \\
SCS deep & $A_{\rm small}$ & $7.547\times10^{-5}$ & 10/10 & 27.5 & unresolved & ($0.0009\%$) \\
SCS deep & $A_{\rm mod}$ & $3.963\times10^{-4}$ & 7/10\textsuperscript{a} & 15.2 & unresolved & ($0.0006\%$) \\
SCS shelf & $A_{\rm mod}$ & $2.354\times10^{-3}$ & 10/10\textsuperscript{b} & 6.8 & $0.30\%/0.30\%$ & $0.027\%$ ($0.030\%$) \\
\bottomrule
\end{tabular}

\vspace{4pt}
\begin{minipage}{\linewidth}
\footnotesize
\textsuperscript{a}The Ritz core, which fails only the $r_\perp$ bar
of items 3, 6 and 10, at the thin surface pycnocline
(\S\ref{sec:mode2:scs}).\\
\textsuperscript{b}Assessed item by item against the same definitions
and bars, without meeting the stricter gradient tolerance of the
stationarity test.\\
The cross-amplitude ordering of item 9 is unmeasured, as for every
single-amplitude configuration.
\end{minipage}
\end{table}

\subsection{What controls the radiation}
\label{sec:mode2:control}

\begin{figure}
  \centerline{\figpdf{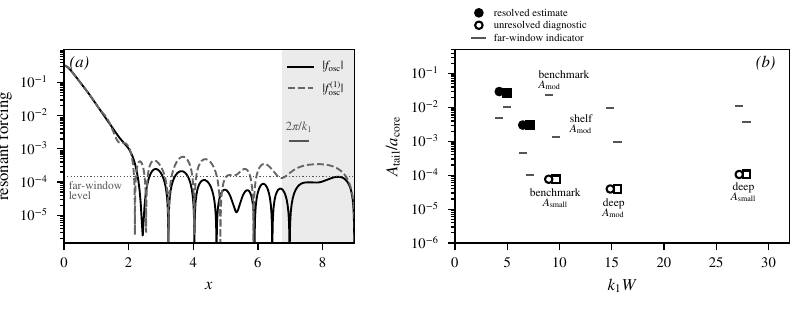}}
  \caption{Linear radiation diagnostics. ($a$)~Magnitude of the
  full-residual forcing $f_{\mathrm{osc}}$ and of its multiplier part
  $f^{(1)}_{\mathrm{osc}}=-f_1\langle w,\psi_1\rangle$ for the moderate
  benchmark core on its base domain, $L=9$. The forcing decays from the
  core to the level of the far window (shaded), which sets the
  far-window indicator, and the resonant wavelength $2\upi/k_1$ is
  marked. ($b$)~Peak outgoing displacement divided by core amplitude against
  $k_1W$, the resonant wavenumber times the core half-width, on base
  (circles) and extended (squares) domains, offset slightly for
  visibility. Filled markers are numerically resolved estimates within
  the approximation, and open markers are unresolved diagnostics, not
  bounds. Grey bars show the far-window indicator, which does not
  include all core uncertainty. The deep moderate core is its Ritz
  state.}
  \label{fig:tailmech}
\end{figure}

The two resolved estimates are too few to establish a scaling law, but
they point to the relevant parameter. The shelf core is the larger of
the two in scaled amplitude, $0.086H$ against $0.046H$, so its forcing,
which is of second order in amplitude, is not the weaker. Yet its tail
relative to its amplitude is about ten times smaller. What
distinguishes the two is the core width measured in resonant
wavelengths, $W/\lambda_1=k_1W/2\upi$, which is $1.09$ on the shelf
against $0.73$ for the benchmark core, that is $k_1W=6.8$
against $4.6$ (figure~\ref{fig:tailmech}$b$). The physical
picture is that of a forced oscillator. The core is a localized source
moving at the phase speed of the mode-1 wave it excites, and only its
Fourier component at the resonant wavenumber $k_1$ radiates
\citep{akylas1992}. A core several resonant wavelengths wide has almost
no content at $k_1$, whereas a core narrower than one wavelength projects
strongly onto it, although a rigid $\mathrm{sech}^2$ core would predict a
contrast nearer a $100\times$ than the observed $10\times$. The
unresolved cores, all with $k_1W\ge9$, neither confirm nor contradict
the trend. Their diagnostics lie below far-window indicators of
$0.1\%$--$2.3\%$ of the core amplitude. In the symmetric sector the forcing vanishes identically, since
an antisymmetric residual has no projection onto the symmetric resonant
mode.

\subsection{Comparison with South China Sea observations}
\label{sec:mode2:roster}

At a mooring near the
$350$~m isobath on the northern South China Sea slope, \citet{yang2009}
record mode-2 solitary waves with isopycnal displacements of
$20\pm14$~m at $75$~m depth and $22\pm15$~m at $240$~m in summer, and
$30\pm18$~m and $26\pm16$~m in winter, fitted by $\mathrm{sech}^2$
profiles with time scales of $8.0\pm4.3$~min and $6.9\pm4.6$~min. The
computed shelf core, on a column of the same depth, places its
elevation lobe at $68$~m and its depression lobe at $230$~m, close to
the observed $75$~m and $240$~m, with displacements of $21$~m and $30$~m
inside the observed ranges and the observed sign pattern, upper
isopycnals raised and lower ones depressed. Its half-width crossing
time is $W/c=10.4$~min and its $\mathrm{sech}^2$-equivalent time
$11.7$~min, longer than the observed mean time scales but within their
spread. The comparison is qualitative. The APE of the shelf core is
prescribed, not fitted to the observations, so its displacements are
consistent with the broad observed spread rather than predicted by it.
The lobe depths, set by the stratification, are the firmer check. On the annual-mean profile and within the linear
prescribed-core estimate, the shelf core loses about $0.03\%$ of its
energy over 100 half-widths, about $39$~km, so intrinsic radiation is
weak over that distance. This
does not determine the propagation range or the lifetime of the
observed waves, which arrive mainly in winter, and it does not explain
the short life of the New Jersey shelf waves of \citet{shroyer2010},
which live in a sheared, dissipative environment on a different
stratification. The steeper deep-basin core, with a $119$~m displacement
in a $3000$~m column, has no observed counterpart. It is a prediction
of a constrained quasi-steady core with a derived radiation estimate,
not a homogeneous-DJL solution and not a hindcast.

\begin{table}
\centering
\caption{Dimensional properties of the South China Sea mode-2 cores and
their prescribed-profile radiation estimates. The deep configuration is
the WOA-fitted $H=3000$~m profile, and the shelf configuration is the
same observational profile truncated at $350$~m, the isobath of the
mooring of \citet{yang2009}, and renormalized. The deep tail amplitudes
are unresolved. The conditional loss estimates assume that unobserved
resonant source errors are no larger than the aggregate assessed
sensitivities and are not rigorous bounds.}
\label{tab:scs}
\small
\begin{tabular}{@{}P{4.2cm}P{2.6cm}P{2.6cm}P{2.5cm}@{}}
\toprule
 & Deep $A_{\rm small}$ & Deep $A_{\rm mod}$ & Shelf $A_{\rm mod}$ \\
\midrule
Depth $H$ (m) & 3000 & 3000 & 350 \\
Wave speed $c$ (m\,s$^{-1}$) & 1.4388 & 1.4601 & 0.6225 \\
Linear reference speeds $(c_{\rm lw2},c_{\rm lw1})$ (m\,s$^{-1}$)
 & (1.4265, 2.7020) & (1.4265, 2.7020) & (0.5991, 1.2780) \\
Maximum isopycnal displacement (m) & 41.8 & 119.5 & 29.9 \\
Half-width (km) & 7.15 & 4.09 & 0.387 \\
Node depth (m) & 423 & 414 & 132 \\
Extremum or lobe depths (m) & 1189 (crest) & 1154 (crest) & 230 / 68 \\
Mode-2 polarity ($\mathrm{sign}\,r_{10,2}$) & concave ($-$) &
concave ($-$) & convex ($+$) \\
Mode-1 peak tail estimate & unresolved & unresolved &
9.1 cm (0.30\% of core) \\
Resonant wavelength (km) & 1.63 & 1.69 & 0.355 \\
Estimated energy loss over 100 half-widths & --- & --- & $0.027\%$ \\
Conditional upper loss estimate over 100 half-widths
 & $0.0009\%$ & $0.0006\%$ & $0.030\%$ \\
\bottomrule
\end{tabular}
\end{table}

\section{Discussion}\label{sec:discussion}

\subsection{What has been established}
\label{sec:discussion:established}

A constrained deep-learning framework is developed that combines variational wave selection with residual refinement to compute internal solitary waves governed by the DJL equation. Turkington's principle selects the
solitary wave against broadening because the APE is a nonlinear
functional of displacement, and the exact factorization
\eqref{eq:broadlimit} says precisely how. The energy partition
$\tfrac12(1+A/E_K)$ favours broadening at every amplitude, since a
localized wave always carries more kinetic energy than APE. The
APE enhancement, which arises because displaced pycnocline fluid
retains the work done against the strong stratification it has
crossed, reverses the ordering and makes the wave the minimizer within
the tested deformation families. Both factors tend to unity in the
weakly nonlinear limit, so selection against broadening weakens as
waves become small and broad, and the local width preference is weak at
every amplitude.

Selection and resolution fall to different objectives. From one
consistently normalized KdV-derived start in one trial space, the
constrained Deep Ritz objective recovers the wave and its multiplier
converges to $1/c^2$, whereas both residual objectives lose the
prescribed APE in the first descent round. The trivial field is a
stationary point of each augmented Lagrangian. The
Rayleigh-initialized Ritz objective is stationary along the amplitude
direction through the start while the residual objectives descend
steeply along it. At the initial multiplier and penalty values, collapse substantially reduces either residual-based objective, whereas it increases the Ritz objective relative to the initial field. These
properties are consistent with the outcome without isolating its cause.
A small residual is therefore insufficient evidence of a wave, and for
residual objectives exact constraint control, physical structure and
independent evaluation are essential. Conversely, the Ritz residual is
dominated by high-wavenumber error that the residual weights by the
fourth power of wavenumber, and a constrained residual refinement of
the recovered wave changes the field by $0.05\%$ while meeting every
criterion, the improvement coming from the spectral content of the
error rather than its size.

Structural exclusion of mode 1 makes asymmetric mode-2 cores
accessible, and the verified machinery produces new solutions, seven
mode-2 waves and cores on four stratifications ending at an
observation-fitted South China Sea profile. On a $350$~m shelf, the
depth of the mooring of \citet{yang2009}, the computed core reproduces
the observed vertical structure qualitatively, with lobes at $68$~m and
$230$~m against the observed $75$~m and $240$~m and displacements inside
the broad observed ranges, and it passes all ten verification items. On
the deep column the steepest core, the state the selection principle
picks, misses the residual criterion at the thin surface pycnocline,
and lowering its residual by refinement moves it along the nearly flat
width direction of the energy, the division of labour found for mode 1.
In the linear prescribed-core estimate, the moderate benchmark core,
narrower than one resonant wavelength at $W/\lambda_1\approx0.73$, loses several per cent
of its energy over 100 half-widths, whereas the shelf core, at
$W/\lambda_1=1.09$ and despite its larger scaled amplitude, loses about
$0.03\%$. The two resolved estimates suggest that radiation weakens as
the core widens relative to the resonant mode-1 wavelength, as a forced
resonance would, and the unresolved cores neither confirm nor contradict
this. Weak intrinsic
radiation bears on, but does not establish, the persistence of observed
mode-2 features.

For individual mode-1 waves, the conventional solvers \citep{dunphy2011,turkington1991} remain faster
and more accurate than the neural calculations
(table~\ref{tab:benchmarks}). The practical advantage of the neural formulation lies in how readily
constrained trial fields are built and optimized. The representation
enforces the boundary conditions and horizontal symmetry exactly, while
antisymmetry or the modal projection restricts the admissible vertical
structure, and this is what makes the mode-2 computations possible. The
same representation serves both the constrained energy minimization
that selects the wave and the residual refinement that resolves it, so
the objective can change without rebuilding the field approximation.
Automatic differentiation carries derivatives through the APE
normalization, the Rayleigh speed and the modal projection, and the
field can be evaluated on independent grids for verification. Together
these features give one implementation for mode-1 waves and constrained
mode-2 cores. Classical methods can impose the same restrictions through
suitable adaptations of their trial spaces.
Accuracy and physical consistency are assessed using the
boundary-corrected virial identity, the joint acceptance criteria,
independent spectral references where available, and grid and domain
checks. An independent repeat of the mode-1 computations reproduces
the energy-landscape results, the contrasting outcomes of the Ritz
and strong objectives from the shared start, the residual refinement,
and wave recovery from the successful random starts.

\subsection{Limitations}
\label{sec:discussion:open}

The quasi-steady cores are not homogeneous-DJL solutions and lack an
independent spectral reference. Their assessment relies on projected
residuals, virial consistency and grid and domain checks. The deep
moderate core fails the orthogonal-remainder criterion, and the
refinement that meets it widens the core by $19\%$. The radiation estimate
neglects finite-amplitude feedback and shared core and truncation
errors, neither of which is bounded by agreement between domains, and
it is not validated against time-dependent simulation, which is the
natural next test. The tails of the deep-water and small benchmark cores are
unresolved, and the weak losses there are conditional. The amplitude dependence
of the mode-1 residuals is unresolved against run-to-run scatter, and
all South China Sea results refer to an annual-mean profile. Profiles
indistinguishable at KdV order are known to have differing fully
nonlinear fates, so the dependence on stratification is a genuine open
question.

\subsection{Outlook}
\label{sec:discussion:outlook}

The structural flexibility of neural DJL solvers provides a basis for
extending these computations to more general physical settings.
Background shear $U(z)$ shifts the linear speed windows and therefore
changes the mode-2 embedding and radiation mechanism discussed in
\S\ref{sec:mode2}. Extending the variational approach to this setting
requires reformulating Turkington's principle, building on the fully
nonlinear formulations with background current of \citet{stastna2002}.
Incorporating topography requires reconsidering the moving-frame
formulation, with classical mapped-grid methods providing an alternative
numerical approach. Topographically trapped waves must also be
distinguished from the trapped cores within travelling waves studied
by \citet{helfrich2010}. Another direction is to represent solution
families using a single conditioned network. Operator learning, through
DeepONet \citep{lu2021deeponet} or Fourier neural operators
\citep{li2021fno}, could map stratification, APE and a specified solution
branch to the wave field and speed. The branch must be specified because
stratification and APE alone need not distinguish coexisting solutions
\citep{dunphy2011}.
Finally, verified DJL fields could supply physically consistent initial
and boundary data for data-assimilative and inverse studies using remote
sensing and mooring observations, as well as for models of wave--structure
interaction and sediment transport.

The closing point returns to where the paper began. The DJL equation is
a nonlinear eigenvalue problem, and for such problems an optimization
objective is not a neutral measure of fit but a statement of which
solution deserves to exist. In the computations reported here,
selection follows the presence of a variational principle in the
objective and accuracy follows how directly the objective controls the
residual. Before adopting a residual objective for a problem of this
kind, one should ask what selects the solution, and if the answer is
nothing, the objective rather than the network is the thing to change.

\begin{bmhead}[Funding.]
This work was supported by the National Natural Science Foundation of
China (42676017), National Key R\&D Program of China (2025YFF0525404), Guangdong Basic and Applied Basic Research Foundation (2025A1515011827)
\end{bmhead}

\begin{bmhead}[Declaration of interests.]
The authors report no conflict of interest.
\end{bmhead}

\begin{bmhead}[Data availability statement.]
The code and data supporting this study are released with the paper,
together with the append-only research record, which contains the
protocols fixed in advance of each stage, the acceptance criteria, the
computational outputs and the full commit history. Two solver toolkits,
one for mode 1 and one for mode 2, together with the scripts that
regenerate every figure and recompute every number in this paper from
the recorded data, are released in the same archive. Every computation is included with its protocol, provenance records, checkpoints, logs and predictions,
including the two predictions that are falsified.
\end{bmhead}

\appendix
\renewcommand{\theHsection}{appendix.\Alph{section}}
\section{Numerical details}\label{app:numerics}

The Ritz computations and the residual comparison share the protocol summarized in table~\ref{tab:protocol}. The exploratory strong-form calculations use mixed collocation sampling, soft boundary
penalties and different optimization and constraint schedules.
All arithmetic, including the scalar returned to the line search, is in
double precision. Functionals and multiplier updates use a fixed
endpoint-trapezoid grid of at least $512\times256$ nodes. For the
largest-amplitude reference wave, $\mathcal{K}$ at this resolution agrees
with its $2048\times1024$ value to a relative $5\times10^{-11}$.
Optimization proceeds in
stages. Stage W fits the initial field in the least-squares sense, by
Adam and then L-BFGS, with separate checks on the fit and on the APE.
Stages R and P are L-BFGS rounds and restarts at fixed
$(\lambda,\gamma)$, with the multiplier, Rayleigh speed and, if the
constraint violation fails to contract, the penalty updated between
them. The mode-2 routines use profile-relative speed and domain checks,
record polarity rather than rejecting reversals, and add an
energy-consistency check between optimization and refined grids,
embeddedness $c<\clw[1]$ and a structure check. 

\begin{table}
\centering
\caption{Main protocol constants for the Deep Ritz computations and the residual comparison, with departures stated in the text. Here
$A_{\rm ref}=0.029333$ and $L=8$. The mode-2 penalty and iteration-count changes
are listed in the last two rows.}
\label{tab:protocol}
\footnotesize
\begin{tabular}{@{}P{2.6cm}P{6.0cm}P{4.0cm}@{}}
\toprule
Item & Value & Note \\
\midrule
Network & tanh MLP, width 64, depth 5, float64 & 16,897 parameters \\
Initialization & Xavier-normal, tanh gain & two network initializations reported \\
Optimization grid & $512\times256$, endpoint trapezoid, float64 & no sampling \\
Validation grids & $500\times250$ headline. $1024\times512$ midpoint & never the optimization grid \\
Stage W & Adam, 4000 iterations at learning rate $3\times10^{-3}$, then L-BFGS, 3000 iterations & least-squares fit only \\
Stage W acceptance & MSE $\le 10^{-6}(A/A_{\rm ref})$ and $|F-A|/A < 10^{-2}$ & MSE threshold scales with $A$ \\
Stage R & 6 rounds $\times$ L-BFGS, \texttt{max\_iter} 800 & monotone descent \\
Stage P & 3 restarts $\times$ L-BFGS, \texttt{max\_iter} 800 & multiplier update between \\
L-BFGS settings & \texttt{strong\_wolfe}, lr 1.0, history 100 & \texttt{tol\_grad} $10^{-16}$, \texttt{tol\_change} $10^{-18}$ \\
Multiplier initialization & $\lambda_0 = 1/c_0^2$ (Rayleigh at the fitted field) & zero for the two tests in \S\ref{sec:verification:mechanism} \\
Penalty $\gamma_0$ & $200\,(A_{\rm ref}/A)$ & $\gamma_0 A$ held constant \\
Penalty escalation & $\times 2$, cap $10^4$, on failure to contract $|F-A|$ by $\tfrac12$ & \\
Residual iteration stopping test & $|F-A|/A < 10^{-6}$ and $|\Delta\mathcal{L}| < 10^{-11}$ between successive line-search evaluations & not a stationarity test. Otherwise iteration limit \\
\midrule
Supercriticality check & $c \ge 1.01\,\clw$ & profile-relative threshold for mode 2 \\
Domain-adequacy check & $\mu_{\rm tail}(c)\,L \ge 8$ & re-checked when $c$ moves \\
Crest-abscissa check & $|x_{\rm crest}| \le 0.05$ & \\
Amplitude check & $\max|\eta| \ge 0.5\times$ initial field & \\
Crest-drift check & $|z_{\rm crest} - z_{\rm start}| \le 0.25$ & wider than the $\pm0.05$ criterion \\
\midrule
Mode-2 penalty & $\gamma_0 = \lambda_0/A$ (3 s.f.), $\gamma_{\max} = 100\gamma_0$ & re-derived, not transplanted \\
Mode-2 stages & W with Adam 2000 and L-BFGS 1500, R $6\times800$, P $3\times800$ & the small benchmark core fits with Adam 8000 and L-BFGS 6000 \\
\bottomrule
\end{tabular}
\end{table}

The residual refinement of \S\ref{sec:mode1:polish} starts from the
saved Ritz endpoint. Writing $\eta=s(\theta)u_\theta$, where $s>0$
solves $\mathcal{F}[su_\theta]=A$ on the original quadrature, the
implicit derivative of the normalization is
\[
 \partial_\theta s=-s\,
 \frac{\iint N^2(z-\eta)\eta\,\partial_\theta u_\theta}
      {\iint N^2(z-\eta)\eta\,u_\theta} .
\]
The objective is $\int_\Omega r^2$, with both the normalization and the
Rayleigh speed differentiated, and the complete gradient is checked by
directional finite differences before optimization. L-BFGS retains the
settings of table~\ref{tab:protocol}, with one iteration and at most 25
evaluations per call and a history persisting across calls. A failed
physical check stops the calculation, as does joint acceptance on the
$1024\times512$ and $2048\times1024$ grids within a fixed 60-minute
limit. The reference enters only the evaluation. For the deep moderate
mode-2 core the objective is the mean-square $r_\perp$, checked every
five steps and stopped once $r_\perp\le3\times10^{-2}$ or the field has
changed by more than $10\%$. That core is optimized on a 512-point
vertical grid, the other mode-2 cores on 256 points.

For the projected mode-2 ansatz \eqref{eq:projansatz}, the projection
quadrature order is computed per stratification. A leakage scan fixed
in advance sets it at 192 for the deep profile, and it is 96 for the shelf, where the measured leakage is $3\times10^{-10}$. The deep and shelf extended domains are reached by adding
an optimized correction to the zero-extended base field, followed by APE
rescaling. Such a transfer can retain a first-derivative jump at the
former boundary, a junction term that the radiation integration-by-parts
check retains. The moderate
benchmark core is instead refitted onto a fresh $L=14$ network from its
zero-extended $L=9$ field and reoptimized on a $512\times256$ grid, so
its horizontal spacing increased from $9/511$ to $14/511$.

A radiation estimate is classified as resolved when, on both base and
extended domains $d=b,e$, the modal amplitude exceeds the indicator
$n_d=4L_d\sigma_{\mathrm{far},d}/(\upi k_{1,d})$, where
$\sigma_{\mathrm{far},d}$ is the r.m.s. of $f_{\mathrm{osc}}$ over the
last quarter of the domain, and the two amplitudes agree within a
factor of two. Otherwise it is unresolved and is not an upper bound.
For the aggregate sensitivity, with peak-positive unit-$L^2$ resonant
modes, the signed overlap
$I_p=\int_0^{L_p}\langle r,\psi_{1,p}\rangle\cos(k_{1,p}x)\,\mathrm{d}x$,
$C_p=2\max|\psi_{1,p}|/k_{1,p}$, $S_p=C_pI_p$, $A_p=|S_p|$ and
$R_p=C_p|I_{\perp,p}|$, with $I_{\perp,p}$ evaluated using $r_\perp$,
the conditional estimate of the peak tail is
\[
 A_{\mathrm{sens}}=\max_p(A_p+R_p)+\operatorname{range}_pS_p
   +\max_d(B_d+T_d)+\max_d\delta_{\mathrm{num},d} ,
\]
where $p$ ranges over the two domains and any saved reproducibility or
wider-network profiles, $B_d$ and $T_d$ are the outer-boundary and
transfer-junction contributions, and $\delta_{\mathrm{num},d}$ sums the
last eigenfunction-grid change, the larger Galerkin or
Galerkin/finite-difference difference and the direct versus
integration-by-parts mismatch. The specification, profile lists and implementation are included in the data archive. No aggregate estimate is
assigned to the small benchmark core. This empirical prescription does
not calibrate a confidence interval or bound shared errors and
neglected feedback.

Costs differ by orders of magnitude (table~\ref{tab:benchmarks}), and the timings carry their own scopes. The spectral reference
solves a mode-1 wave in about $1.9$~s on a $256^2$ grid at its default tolerance, whereas tight-tolerance references on a $512^2$
grid take $63$--$80$~s. The finite-difference Turkington hybrid takes
about $22$~s for 30 iterations. The Deep Ritz computations from the consistent start take about
$2.5\times10^3$~s per wave at the iteration counts used here, three
orders of magnitude above the spectral cost, and placing a mesh-free solver inside the classical outer iteration is not pursued because of a measured $91$-fold per-iteration overhead on a
contended host.  
\begin{table}
\centering
\caption{Mode-1 benchmarks on the Dunphy problem at $A_{\rm ref}$.
Procedures and grid conventions differ between rows. The Ritz
rows from the consistent start of \S\ref{sec:formulations:kdv} use two
network initializations, and the Ritz endpoint and its residual
refinement are separate stages, the refinement time excluding the
Ritz recovery and the initial fitting. Speed accuracy does not
substitute for field or residual accuracy.}
\label{tab:benchmarks}
\small
\begin{tabular}{@{}P{3.1cm}P{2.5cm}P{1.8cm}P{1.5cm}P{2.5cm}@{}}
\toprule
Solver & res\textsubscript{rms} & $c$ rel.\ err. & Elapsed time & Note \\
\midrule
PyDJL (spectral reference) & $4.4\times10^{-8}$
 & (reference) & $\sim$1.9 s & reference solution \\
FD-inner Turkington hybrid & $2.51\times10^{-4}$ & $2.22\times10^{-5}$
 & $\sim$22.3 s & classical comparison \\
Strong-form PINN & $8.7\times10^{-4}$ & $1.78\times10^{-5}$
 & --- & reference-initialized \\
VPINN (512 tests) & $1.07\times10^{-2}$
 & $1.86\times10^{-6}$ & --- & reference-initialized \\
Deep Ritz (from reference) & $1.208\times10^{-2}$ & --- & --- &
measured residual \\
\midrule
Deep Ritz (consistent start) & $9.679\times10^{-3}$/$1.16\times10^{-2}$
 & $(2.1/4.1)\times10^{-6}$ & --- &
wave recovered, max.\ field error $(4.6/7.0)\times10^{-4}$. Residual fails \\
Residual refinement & $9.986\times10^{-4}$ & $2.326\times10^{-6}$
 & 36.38 min & joint criteria pass on both grids \\
\bottomrule
\end{tabular}
\end{table}

\section{Search with a deflated functional}\label{app:uniqueness}

The search of \S\ref{sec:mode1:uniqueness} multiplies the scalar
augmented Lagrangian $J$ by $M=\prod_i[1+\sigma/d_i(\theta)^2]$, where
$d_i$ is the full-domain $L^2$ distance from a previously known field
and $\sigma=(0.1R_0)^2$, with $R_0$ the norm of the known wave. Unlike
the residual deflation of \citet{farrell2015}, this multiplication
changes stationary points in general, since
$\nabla_\theta(MJ)=M\nabla_\theta J+J\nabla_\theta M$, so it is a
heuristic search for competing fields rather than a uniqueness test.
The known field is the Ritz wave recovered from the consistent start
(\S\ref{sec:verification:comparison}). Five deterministic starts are
used, namely a mode-2 vertical structure on the wave's half-width, the
wave with its sign reversed, an off-axis displaced pair, the wave
broadened by $\beta=3$ and a two-lobe near-overlap pair, each rescaled to
the prescribed APE. The mode-2 start and the $\beta=3$ start end on
candidate $X_1$, a single core broadened to $1.49$ times the wave's
half-width at a distance $0.25R_0$, and the two pair starts end on
candidate $X_2$, a single core compressed to $0.78$ times the half-width
at $0.22R_0$. Their residuals, $2.2$--$2.7\times10^{-2}$ and
$3.4$--$3.5\times10^{-2}$, and their virial discrepancies,
$6.0\times10^{-2}$ and $8.8\times10^{-2}$, place both well outside the
acceptance criteria. Both are reproduced to four digits on the
$1024\times512$ grid, so they are not quadrature artefacts but shells
held off the wave by the deflation factor, which the weak width
preference of \S\ref{sec:djl:landscape} lets sit at about twice the
nominal deflation scale. The elevation start overturns too strongly
to be fitted within the acceptance criteria of Stage W, and with that
check waived it ends on the near-lid artefact of
\S\ref{sec:mode1:reachability}, whose $\mathcal{K}$ changes eightfold
under grid refinement. No start produces an additional accepted
single-core wave.

\bibliographystyle{jfm}
\bibliography{references}

\begin{thebibliography}{30}
\expandafter\ifx\csname natexlab\endcsname\relax\def\natexlab#1{#1}\fi
\def\au#1{#1} \def\ed#1{#1} \def\yr#1{#1}\def\at#1{#1}\def\jt#1{\textit{#1}}
  \def\bt#1{#1}\def\bvol#1{\textbf{#1}} \def\vol#1{#1} \def\pg#1{#1}
  \def\publ#1{#1}\def\arxiv#1{#1}\def\org#1{#1}\def\st#1{\textit{#1}}

\bibitem[Akylas \& Grimshaw(1992)]{akylas1992}
{\sc \au{Akylas, T.~R.} \& \au{Grimshaw, R. H.~J.}} \yr{1992}  \at{Solitary
  internal waves with oscillatory tails}.  \jt{Journal of Fluid Mechanics}
  \bvol{242},  \pg{279--298}.

\bibitem[Carr {\em et~al.\/}(2015)Carr, Davies \& Hoebers]{carr2015}
{\sc \au{Carr, Magda}, \au{Davies, Peter~A.} \& \au{Hoebers, R.~P.}} \yr{2015}
  \at{Experiments on the structure and stability of mode-2 internal
  solitary-like waves propagating on an offset pycnocline}.  \jt{Physics of
  Fluids}  \bvol{27}~(4),  \pg{046602}.

\bibitem[Davis \& Acrivos(1967)]{davis1967}
{\sc \au{Davis, Russ~E.} \& \au{Acrivos, Andreas}} \yr{1967}  \at{Solitary
  internal waves in deep water}.  \jt{Journal of Fluid Mechanics}
  \bvol{29}~(3),  \pg{593--607}.

\bibitem[Doak {\em et~al.\/}(2022)Doak, Barros \& Milewski]{doak2022}
{\sc \au{Doak, Alex}, \au{Barros, Ricardo} \& \au{Milewski, Paul~A.}} \yr{2022}
   \at{Large mode-2 internal solitary waves in three-layer flows}.  \jt{Journal
  of Fluid Mechanics}  \bvol{953},  \pg{A42}.

\bibitem[Dubreil-Jacotin(1934)]{dubreil1934}
{\sc \au{Dubreil-Jacotin, Marie-Louise}} \yr{1934}  \at{Sur la
  d{\'e}termination rigoureuse des ondes permanentes p{\'e}riodiques d'ampleur
  finie}.  \jt{Journal de Math{\'e}matiques Pures et Appliqu{\'e}es}
  \bvol{13},  \pg{217--291}.

\bibitem[Dunphy(2011)]{djles}
{\sc \au{Dunphy, Michael}} \yr{2011} {DJLES}: a {MATLAB} solver for the
  {Dubreil-Jacotin-Long} equation. \url{https://github.com/mdunphy/DJLES}.

\bibitem[Dunphy {\em et~al.\/}(2011)Dunphy, Subich \& Stastna]{dunphy2011}
{\sc \au{Dunphy, Michael}, \au{Subich, Christopher} \& \au{Stastna, Marek}}
  \yr{2011}  \at{Spectral methods for internal waves: indistinguishable density
  profiles and double-humped solitary waves}.  \jt{Nonlinear Processes in
  Geophysics}  \bvol{18}~(3),  \pg{351--358}.

\bibitem[E \& Yu(2018)]{eyu2018}
{\sc \au{E, Weinan} \& \au{Yu, Bing}} \yr{2018}  \at{The {D}eep {R}itz method:
  a deep learning-based numerical algorithm for solving variational problems}.
  \jt{Communications in Mathematics and Statistics}  \bvol{6}~(1),  \pg{1--12}.

\bibitem[Farmer {\em et~al.\/}(2011)Farmer, Alford, Lien, Yang, Chang \&
  Li]{farmer2011}
{\sc \au{Farmer, David~M.}, \au{Alford, Matthew~H.}, \au{Lien, Ren-Chieh},
  \au{Yang, Yiing~Jiang}, \au{Chang, Ming-Huei} \& \au{Li, Qiang}} \yr{2011}
  \at{From {L}uzon {S}trait to {D}ongsha {P}lateau: stages in the life of an
  internal wave}.  \jt{Oceanography}  \bvol{24}~(4),  \pg{64--77}.

\bibitem[Farrell {\em et~al.\/}(2015)Farrell, Birkisson \& Funke]{farrell2015}
{\sc \au{Farrell, Patrick~E.}, \au{Birkisson, {\'A}sgeir} \& \au{Funke,
  Simon~W.}} \yr{2015}  \at{Deflation techniques for finding distinct solutions
  of nonlinear partial differential equations}.  \jt{SIAM Journal on Scientific
  Computing}  \bvol{37}~(4),  \pg{A2026--A2045}.

\bibitem[Helfrich \& Melville(2006)]{helfrich2006}
{\sc \au{Helfrich, Karl~R.} \& \au{Melville, W.~Kendall}} \yr{2006}  \at{Long
  nonlinear internal waves}.  \jt{Annual Review of Fluid Mechanics}  \bvol{38},
   \pg{395--425}.

\bibitem[Helfrich \& White(2010)]{helfrich2010}
{\sc \au{Helfrich, Karl~R.} \& \au{White, Brian~L.}} \yr{2010}  \at{A model for
  large-amplitude internal solitary waves with trapped cores}.  \jt{Nonlinear
  Processes in Geophysics}  \bvol{17}~(4),  \pg{303--318}.

\bibitem[Jin {\em et~al.\/}(2022)Jin, Mattheakis \& Protopapas]{jin2022}
{\sc \au{Jin, Henry}, \au{Mattheakis, Marios} \& \au{Protopapas, Pavlos}}
  \yr{2022} Physics-informed neural networks for quantum eigenvalue problems.
  \bt{In {\em 2022 International Joint Conference on Neural Networks
  (IJCNN)\/}},  \pg{pp. 1--8}.  \publ{IEEE}.

\bibitem[Kharazmi {\em et~al.\/}(2021)Kharazmi, Zhang \&
  Karniadakis]{kharazmi2021}
{\sc \au{Kharazmi, Ehsan}, \au{Zhang, Zhongqiang} \& \au{Karniadakis,
  George~E.}} \yr{2021}  \at{hp-{VPINN}s: variational physics-informed neural
  networks with domain decomposition}.  \jt{Computer Methods in Applied
  Mechanics and Engineering}  \bvol{374},  \pg{113547}.

\bibitem[Kingma \& Ba(2015)]{kingma2015}
{\sc \au{Kingma, Diederik~P.} \& \au{Ba, Jimmy}} \yr{2015} Adam: a method for
  stochastic optimization.  \bt{In {\em International Conference on Learning
  Representations\/}}.

\bibitem[Lamb \& Nguyen(2009)]{lamb2009}
{\sc \au{Lamb, Kevin~G.} \& \au{Nguyen, Van~Thinh}} \yr{2009}  \at{Calculating
  energy flux in internal solitary waves with an application to reflectance}.
  \jt{J. Phys. Oceanogr.}  \bvol{39}~(3),  \pg{559--580}.

\bibitem[Li \& Farmer(2011)]{li_jpo2011}
{\sc \au{Li, Qiang} \& \au{Farmer, David~M.}} \yr{2011}  \at{The generation and
  evolution of nonlinear internal waves in the deep basin of the {S}outh
  {C}hina {S}ea}.  \jt{Journal of Physical Oceanography}  \bvol{41}~(7),
  \pg{1345--1363}.

\bibitem[Li {\em et~al.\/}(2021)Li, Kovachki, Azizzadenesheli, Liu,
  Bhattacharya, Stuart \& Anandkumar]{li2021fno}
{\sc \au{Li, Zongyi}, \au{Kovachki, Nikola}, \au{Azizzadenesheli, Kamyar},
  \au{Liu, Burigede}, \au{Bhattacharya, Kaushik}, \au{Stuart, Andrew} \&
  \au{Anandkumar, Anima}} \yr{2021} Fourier neural operator for parametric
  partial differential equations.  \bt{In {\em International Conference on
  Learning Representations\/}}.

\bibitem[Liu \& Nocedal(1989)]{liu1989}
{\sc \au{Liu, Dong~C.} \& \au{Nocedal, Jorge}} \yr{1989}  \at{On the limited
  memory {BFGS} method for large scale optimization}.  \jt{Mathematical
  Programming}  \bvol{45},  \pg{503--528}.

\bibitem[Long(1953)]{long1953}
{\sc \au{Long, Robert~R.}} \yr{1953}  \at{Some aspects of the flow of
  stratified fluids: {I}. {A} theoretical investigation}.  \jt{Tellus}
  \bvol{5}~(1),  \pg{42--58}.

\bibitem[Lu {\em et~al.\/}(2021)Lu, Jin, Pang, Zhang \&
  Karniadakis]{lu2021deeponet}
{\sc \au{Lu, Lu}, \au{Jin, Pengzhan}, \au{Pang, Guofei}, \au{Zhang, Zhongqiang}
  \& \au{Karniadakis, George~Em}} \yr{2021}  \at{Learning nonlinear operators
  via {DeepONet} based on the universal approximation theorem of operators}.
  \jt{Nature Machine Intelligence}  \bvol{3},  \pg{218--229}.

\bibitem[Olsthoorn {\em et~al.\/}(2013)Olsthoorn, Baglaenko \&
  Stastna]{olsthoorn2013}
{\sc \au{Olsthoorn, Jason}, \au{Baglaenko, Anton} \& \au{Stastna, Marek}}
  \yr{2013}  \at{Analysis of asymmetries in propagating mode-2 waves}.
  \jt{Nonlinear Processes in Geophysics}  \bvol{20}~(1),  \pg{59--69}.

\bibitem[{PyDJL}(2019)]{pydjl}
{\sc \au{{PyDJL}}} \yr{2019} {PyDJL}: a {Python} translation of {DJLES}.
  \url{https://github.com/golnazir/PyDJL}, commit 9ab2419 (12 July 2019).

\bibitem[Raissi {\em et~al.\/}(2019)Raissi, Perdikaris \&
  Karniadakis]{raissi2019}
{\sc \au{Raissi, Maziar}, \au{Perdikaris, Paris} \& \au{Karniadakis,
  George~E.}} \yr{2019}  \at{Physics-informed neural networks: a deep learning
  framework for solving forward and inverse problems involving nonlinear
  partial differential equations}.  \jt{Journal of Computational Physics}
  \bvol{378},  \pg{686--707}.

\bibitem[Shroyer {\em et~al.\/}(2010)Shroyer, Moum \& Nash]{shroyer2010}
{\sc \au{Shroyer, Emily~L.}, \au{Moum, James~N.} \& \au{Nash, Jonathan~D.}}
  \yr{2010}  \at{Mode 2 waves on the continental shelf: ephemeral components of
  the nonlinear internal wavefield}.  \jt{Journal of Geophysical Research:
  Oceans}  \bvol{115},  \pg{C07001}.

\bibitem[Song {\em et~al.\/}(2024)Song, Zhong, Karniadakis \& Yan]{song2024}
{\sc \au{Song, Jin}, \au{Zhong, Ming}, \au{Karniadakis, George~Em} \& \au{Yan,
  Zhenya}} \yr{2024}  \at{Two-stage initial-value iterative physics-informed
  neural networks for simulating solitary waves of nonlinear wave equations}.
  \jt{J. Comput. Phys.}  \bvol{505},  \pg{112917}.

\bibitem[Stastna \& Lamb(2002)]{stastna2002}
{\sc \au{Stastna, Marek} \& \au{Lamb, Kevin~G.}} \yr{2002}  \at{Large fully
  nonlinear internal solitary waves: the effect of background current}.
  \jt{Physics of Fluids}  \bvol{14}~(9),  \pg{2987--2999}.

\bibitem[Turkington {\em et~al.\/}(1991)Turkington, Eydeland \&
  Wang]{turkington1991}
{\sc \au{Turkington, Bruce}, \au{Eydeland, Alexander} \& \au{Wang, Shouhong}}
  \yr{1991}  \at{A computational method for solitary internal waves in a
  continuously stratified fluid}.  \jt{Studies in Applied Mathematics}
  \bvol{85}~(2),  \pg{93--127}.

\bibitem[Wang {\em et~al.\/}(2021)Wang, Teng \& Perdikaris]{wang2021}
{\sc \au{Wang, Sifan}, \au{Teng, Yujun} \& \au{Perdikaris, Paris}} \yr{2021}
  \at{Understanding and mitigating gradient flow pathologies in
  physics-informed neural networks}.  \jt{SIAM J. Sci. Comput.}  \bvol{43}~(5),
   \pg{A3055--A3081}.

\bibitem[Yang {\em et~al.\/}(2009)Yang, Fang, Chang, Ramp, Kao \&
  Tang]{yang2009}
{\sc \au{Yang, Yiing~Jang}, \au{Fang, Ying~Chih}, \au{Chang, Ming-Huei},
  \au{Ramp, Steven~R.}, \au{Kao, Chih-Chung} \& \au{Tang, Tswen~Yung}}
  \yr{2009}  \at{Observations of second baroclinic mode internal solitary waves
  on the continental slope of the northern {S}outh {C}hina {S}ea}.  \jt{Journal
  of Geophysical Research: Oceans}  \bvol{114},  \pg{C10003}.

\end{thebibliography}
\end{document}